\documentclass[11pt]{article}
\usepackage{mathrsfs}
\usepackage{amsmath}
\usepackage{amsfonts}
\usepackage{amssymb, amsmath, cite}
\usepackage{color}
\usepackage{graphicx,float}
\usepackage{pgfplots}

\renewcommand{\theequation}{\thesection.\arabic{equation}}

\newcommand\be{\begin{equation}}
\newcommand\ee{\end{equation}}
\newcommand\ber{\begin{eqnarray}}
\newcommand\eer{\end{eqnarray}}
\newcommand\berr{\begin{eqnarray*}}
\newcommand\eerr{\end{eqnarray*}}
\newcommand\bea{\begin{eqnarray}}
\newcommand\eea{\end{eqnarray}}

\newcommand{\bfR}{{\Bbb R}}

\newcommand{\x}{{\bf x}}

\newcommand{\dd}{\mbox{d}}
\newcommand{\e}{\mbox{e}}
\newcommand{\pa}{\partial}\newcommand{\Om}{\Omega}

\newcommand{\nn}{\nonumber}

\newcommand\lb{\label}
\newcommand\eq{\eqref}

\newcommand{\vp}{\varphi}

\title{Dyonic Matter Equations, Exact Point-Source Solutions, and Charged Black Holes in Generalized Born--Infeld Theory}

\author{
Yisong Yang\footnote{Email address: yisongyang@nyu.edu}\\Courant Institute of Mathematical Sciences\\ New York University\\New York, New York 10012, USA
}

\date{}

\begin{document}

\maketitle

\begin{abstract}
We derive  the equations of motion governing static dyonic matters, described in terms of two real scalar fields, 
in nonlinear electrodynamics of the Born--Infeld theory type. We then obtain exact finite-energy solutions of these equations in the quadratic and logarithmic nonlinearity cases subject to dyonic point-charge sources
and construct dyonically charged black holes with relegated curvature singularities. In the case of quadratic nonlinearity, which is the core model of this work,  we show that dyonic solutions enable us to
restore electromagnetic symmetry,  which is known to be broken in non-dyonic situations by exclusion of monopoles.
We further demonstrate that in the context of k-essence cosmology the nonlinear electrodynamics models possess their own distinctive signatures in light of the underlying equations of state
of the cosmic fluids they represent. In this context, the quadratic and logarithmic models are shown to resolve a density-pressure inconsistency issue exhibited by the original Born--Infeld model
k-essence action function as well as by all of its fractional-powered extensions. Moreover, it is shown that the quadratic model is uniquely positioned to give rise to a radiation-dominated era
in the early universe among all the polynomial models and other examples considered.
\medskip

{\flushleft {Keywords:} Nonlinear electrodynamics, dyonic matter equations, exact solutions,  charged black holes, curvature singularities, regular
black holes, Hawking temperature, k-essence cosmology, equation of state, radiation-dominated era
\medskip

\medskip

}
\end{abstract}

\section{Introduction}
\setcounter{equation}{0}

It is well known that the motivation of Born and Infeld  \cite{B1,B2,B3,B4} for the introduction of their nonlinear
electromagnetic field theory
is to overcome the energy divergence problem associated with a  point charge source
in the original Maxwell theory,  in order to model the electron as a point charge. In its full formalism \cite{B4}, the Born--Infeld theory contains two related models, one based on a
consideration of the action principle of special relativity, leading to the Lagrangian action density function
\be\lb{1}
{\cal L}=b^2\left(1-\sqrt{1-\frac1{b^2}({\bf E}^2 -{\bf B}^2)}\right),
\ee
where ${\bf E}$ and $\bf B$ are electric and magnetic fields, respectively, and $b>0$ a coupling parameter, and the other on an invariance principle consideration, giving rise to the Lagrangian action density function,
\be\lb{2}
{\cal L}=b^2\left(1-\sqrt{1-\frac1{b^2}({\bf E}^2 -{\bf B}^2)-\frac1{b^4}({\bf E}\cdot{\bf B})^2}\right),
\ee
which contains a higher-order mixed interaction term of the electric and magnetic fields. These models are sometimes referred to as the first and second Born--Infeld models,
respectively.  The static equations of motion of \eq{1} accommodate finite-energy solutions representing either an electric or magnetic point charge, or monopole \cite{Curie,Dirac,GO,JT,Pre} as is
commonly known, but {\em not} a dyon \cite{Yang1}, a finite-energy solution representing a point source carrying both electric and magnetic charges \cite{Sch,JZ,PS,W,Ybook}.
In other words, in the sense of supporting a finite-energy point charge, the Born--Infeld model \eq{1} exhibits an electromagnetic symmetry since both electric and magnetic point charges are
permitted by \eq{1}. On the other hand,
although the notion
of monopoles is conceptually important and fruitful in field theory physics \cite{GO,Pre,Ra,Wein}, monopoles themselves have never been
observed in nature or laboratory in isolation, except for some of their simulated condensed-matter-system realizations \cite{Gib,Raj,Yak}. Inspired by the extensive applications of
the idea of the Born--Infeld theory in recent development, ranging from particle physics \cite{BB}, superstring theory \cite{Tsey}, and modified gravity theories  \cite{JHOR},  it would be interesting
to pursue the idea of the Born--Infeld theory to come up with a nonlinear electrodynamics theory that would exclude monopoles but only accommodate electric point charges. In \cite{Yang1},
such a goal is achieved for the general polynomial model of the form
\be\lb{3}
{\cal L}=p_n(s),\quad p_n(s)=s+\sum_{m=2}^n a_m s^m,\quad s=\frac1{2}({\bf E}^2 -{\bf B}^2),
\ee
and it is shown that the model accommodates finite-energy electric point charges but not monopoles for any $n\geq2$. This phenomenon is specifically referred to as electromagnetic asymmetry
\cite{Yang1} associated with the model \eq{3}. 
The model \eq{3} is of broad interest and importance because it may be regarded as to offer the simplest nonlinearity function for the electrodynamics theory
and unlike the model \eq{1} no finite-range truncation is imposed on $s$. Moreover,
by the Stone--Weierstrass density theorem \cite{Stone,Yosida}, any continuous function over a compact interval may uniformly be approximated by polynomials so that
polynomial type nonlinearities are considered more elementary, although we know that
in theory of electromagnetism the range of $s$ is not compact. Subsequently, it is of interest to know whether electromagnetic symmetry may be restored in polynomial models 
for {\em dyonic} point charges. Since dyons are excluded from \eq{1} but accommodated in \eq{2} as shown in \cite{Yang1,Yang2}, we see that the adequate question to ask is whether dyons are
supported in the polynomial model
\be\lb{4}
{\cal L}=p_n(s),\quad p_n(s)=s+\sum_{m=2}^n a_m s^m,\quad s=\frac1{2}({\bf E}^2 -{\bf B}^2)+\frac{\kappa^2}{2}({\bf E}\cdot{\bf B})^2,
\ee
containing a higher-order mixed interaction term spelled out in the second model of Born--Infeld based on the invariance principle given in \eq{2}, where $\kappa$ is a free parameter which may
be set to zero to recover the model \eq{3} as its limiting situation. A main result of this work is to confirm this inquiry by showing that, in the quadratic situation with $n=2$, the model \eq{4}
 accommodates finite-energy dyons if and only if $\kappa>0$. 
Thus a broken electromagnetic symmetry may be {\em restored} at the dyonic theory level when both electricity and magnetism are present.
Furthermore, an important application of finite-energy electromagnetism is to systematically obtain \cite{Yang1,Yang2} charged black holes with relegated or sometimes
removed singularities at the centers of charge sources as demonstrated earlier in the contexts of the Bardeen \cite{AG1,AG2} and Hayward \cite{BP,Hay,Kumar} magnetically charged black holes, respectively.  The interest and importance of dyons explored thus prompt us to carry out a systematic  derivation of the static dyonic matter equations in the generalized
Born--Infeld theory in terms of a pair of electric and magnetic scalar potentials. Although these equations in a general setup are rather complicated, for some
interesting specific cases we are able to obtain
their exact solutions subject to point-charge sources in explicit forms. These include the classical Born--Infeld model and
the exponential model considered in \cite{Yang2}. In this work, we obtain such dyonic solutions for
the quadratic and logarithmic models as well. It will also be shown that these new solutions give rise to dyonically charged black holes with relegated or
ameliorated curvature singularities as in \cite{Yang2}. That is, they give rise to electrically {\em and} magnetic charged black holes with reduced curvature singularities.
The dyonic solutions of the quadratic model are of particular interest because the model does not allow finite-energy monopole solutions and thus exhibits electromagnetic asymmetry as
described earlier and dyonic solutions with $\kappa>0$ serve to restore electromagnetic symmetry.
 Moreover,  we will see that, in the context of k-essence cosmology,  various generalized Born--Infeld models 
 may further be differentiated with refined cosmic fluid characteristics and dynamics. A notable
feature worth mentioning here is that, in this context, there is a density-pressure inconsistency issue associated with the classical Born--Infeld action function in that, at the big-bang
moment, the k-essence fluid density is infinite but the pressure remains finite. This property clearly violates the general consensus that the early universe is a radiation-dominated
era, for example. We will show that both the quadratic and logarithmic models resolve such an inconsistency issue so that the fluid density
and pressure both become infinite at the big-bang moment and that the equation of state of the quadratic model also gives rise to a radiation-dominated era in the early universe correctly,
and distinctively.

 In Sections 2--3, we schematically derive the dyonic matter equations for the generalized Born--Infeld electrodynamics and present some existence and nonexistence results. 
In Sections 4--5, we construct solutions to the dyonic matter equations for the quadratic and logarithmic models, respectively, and show how the coupling parameter $\kappa$
serves the role to switch on and off the finiteness of energy of the dyonic solutions in the models. In Section 6, we construct charged black holes with relegated curvature singularities in the quadratic and logarithmic models and explain that in these models electromagnetism contributes to black-hole thermodynamics. In Section 7, we characterize the nonlinear dynamics
 models considered in the context of k-essence cosmology. In particular, we show that the quadratic model demonstrates a unique signature that it gives rise to
a radiation-dominated era in the early universe in contrast against all other models considered. In Section 8, we draw conclusions.

\section{Generalized Born--Infeld electromagnetism and dyonic matter equations in the classical model}

\setcounter{equation}{0}

 In this section, we consider the setup of the generalized Born--Infeld nonlinear electrodynamics theory and 
 develop  methods for deriving the associated dyonic matter equations and obtaining their solutions schematically.

Consider the Lagrangian action density \cite{Yang2}
\be\lb{2.8}
{\cal L}=f(s), \quad s=\frac12({\bf E}^2-{\bf B}^2)+\frac{\kappa^2}2({\bf E}\cdot{\bf B})^2.
\ee
giving rise to a generalized nonlinear electrodynamics theory of the Born--Infeld type,
where the action density profile function $f$ is assumed to satisfy the condition
$
f(0)=0, f'(0)=1$, and $\bf E$ and $\bf B$ are electric and magnetic fields, respectively.
Let the associated electric displacement and magnetic intensity fields be denoted by $\bf D$ and $\bf H$, respectively. Then they are related to $\bf E$ and $\bf B$ by
the nonlinear constitutive equations 
\bea
{\bf D}&=&f'(s)\left({\bf E}+{\kappa^2}({\bf E}\cdot{\bf B})\,{\bf B}\right), \lb{2.18} \\
{\bf H}&=&f'(s)\left({\bf B}-{\kappa^2}({\bf E}\cdot{\bf B})\,{\bf E}\right).\lb{2.19}
\eea
This nonlinear theory of electromagnetism may be viewed as 
describing the interaction of electromagnetic fields in a nonlinear medium with field-dependent dielectrics and permeability coefficients of a mixed type. In fact, from \eq{2.18} and \eq{2.19}, we have
\be
\left(\begin{array}{c}{\bf D}\\{\bf B}\end{array}\right)=\Sigma({\bf E},{\bf B})\left(\begin{array}{c}{\bf E}\\{\bf H}\end{array}\right),
\quad \Sigma({\bf E},{\bf B})\equiv\left(\begin{array}{cc}f'(s)(1+\kappa^4({\bf E}\cdot{\bf B})^2)& \kappa^2({\bf E}\cdot{\bf B})\\ \kappa^2({\bf E}\cdot{\bf B})&\frac1{f'(s)}\end{array}\right),
\ee
where the matrix $\Sigma({\bf E},{\bf B})$ contains the dielectrics and permeability information of the system such that the property $\det(\Sigma({\bf E},{\bf B}))=1$ resembles the
constraint that the speed of light in vacuum is normalized to unity.
These relations blend the electric  and magnetic interactions in the generalized Born--Infeld electrodynamics and comprise the main core of all technical difficulties that come along.
For this theory, the Hamiltonian energy density may be calculated to be
\be\lb{2.21}
{\cal H}=
f'(s)\left({\bf E}^2+\kappa^2[{\bf E}\cdot{\bf B}]^2\right)
-f(s).
\ee
This expression will be useful when we evaluate the energy of a dyonically charged source and compute the gravitational metric factor of a charged black hole.

To illustrate the method for deriving the dyonic matter equations in terms of associated scalar potentials, we begin by considering the classical Born--Infeld model defined by
\be\lb{x2.6}
f(s)=\frac1\beta\left(1-\sqrt{1-2\beta s}\right),
\ee
where $\beta>0$ is a coupling parameter independent of the parameter $\kappa$ in \eq{2.8}. This is a relatively simpler situation to deal with. However,
the insight developed will become useful later for us to treat general situations as we will soon see.
For this purpose, we first obtain from \eq{2.19} the 
general expressions
\bea
{\bf E}\cdot{\bf B}&=&\frac{{\bf E}\cdot{\bf H}}{f'(s)(1-\kappa^2{\bf E}^2)},\lb{2.23}\\
{\bf B}&=&\frac{\bf H}{f'(s)}+\kappa^2 ({\bf E}\cdot{\bf B}){\bf E}=\frac1{f'(s)}\left({\bf H}+\frac{\kappa^2({\bf E}\cdot{\bf H})}{1-\kappa^2{\bf E}^2}{\bf E}\right).\lb{2.24}
\eea
Now, specializing on \eq{x2.6}, we can  insert \eq{2.23} and \eq{2.24} into
\be\lb{2.25}
\frac1{f'(s)}\equiv {\cal R}=\sqrt{1-2\beta s}=\sqrt{1-\beta({\bf E}^2-{\bf B}^2+{\kappa^2}[{\bf E}\cdot{\bf B}]^2)}
\ee
to obtain
\be\lb{2.26}
{\cal R}^2=1-\beta{\bf E}^2+\beta{\cal R}^2\left({\bf H}+\frac{\kappa^2 ({\bf E}\cdot{\bf H})}{1-\kappa^2{\bf E}^2}\,{\bf E}\right)^2-\beta\kappa^2{\cal R}^2\left(\frac{{\bf E}\cdot{\bf H}}{1-\kappa^2{\bf E}^2}\right)^2,
\ee
resulting in the solution
\be\lb{2.27}
{\cal R}^2=\frac{1-\beta {\bf E}^2}{1-\beta \left({\bf H}+\frac{\kappa^2({\bf E}\cdot{\bf H})}{1-\kappa^2{\bf E}^2}\,{\bf E}\right)^2+\beta\kappa^2\left(\frac{{\bf E}\cdot{\bf H}}{1-\kappa^2{\bf E}^2}\right)^2}.
\ee
Therefore, in view of \eq{2.18}, \eq{2.19}, \eq{2.23}--\eq{2.25}, and \eq{2.27}, we get
\bea
{\bf D}&=&\frac{\bf E}{{\cal R}}+\kappa^2{\cal R}\frac{({\bf E}\cdot{\bf H})}{1-\kappa^2{\bf E}^2 }\left({\bf H}+\frac{\kappa^2({\bf E}\cdot{\bf H})}{1-\kappa^2{\bf E}^2}\,{\bf E}\right),\lb{2.28}\\
{\bf B}&=&{\cal R}\left({\bf H}+\frac{\kappa^2({\bf E}\cdot{\bf H})}{1-\kappa^2{\bf E}^2}{\bf E}\right).\lb{2.29}
\eea
Furthermore, assume the presence of a dyonic charge distribution given in terms of an electric charge
density function, $\rho_e$, and a magnetic one, $\rho_m$. Then, in the static situation, the governing equations of the theory \eq{2.8} are of the Maxwell type \cite{Sch,Yang2}:
\bea
&&\nabla\times{\bf E}={\bf0},\quad \nabla\times{\bf H}={\bf0},\lb{x2.14}\\
&& \nabla\cdot{\bf D}=\rho_e,\quad \nabla\cdot{\bf B}=\rho_m.\lb{2.31}
\eea
Resolving \eq{x2.14}, we obtain
\be\lb{2.30}
{\bf E}=\nabla\phi, \quad {\bf H}=\nabla\psi,
\ee
 for some real-valued scalar functions $\phi$ and $\psi$, respectively, i.e., a pair of electric and magnetic scalar potentials.  
Consequently, inserting \eq{2.30} into \eq{2.27}--\eq{2.29},   we see that \eq{2.31} renders the equations
\bea
&&\nabla\cdot\left(\frac{\nabla\phi}{\cal R}+\kappa^2{\cal R}\frac{\nabla\phi\cdot\nabla\psi}{1-\kappa^2|\nabla\phi|^2}\left[\nabla \psi+\frac{\kappa^2(\nabla\phi\cdot\nabla\psi)}{1-\kappa^2|\nabla\phi|^2}\nabla\phi\right]\right)=\rho_e,\lb{2.32}\\
&&\nabla\cdot\left({\cal R}\left[\nabla\psi+\frac{\kappa^2(\nabla\phi\cdot\nabla\psi)}{1-\kappa^2|\nabla\phi|^2}\nabla\phi\right]\right)=\rho_m,\lb{2.33}\\
&&{\cal R}=\sqrt{\frac{1-\beta |\nabla\phi|^2}{1-\beta \left(\nabla\psi+\frac{\kappa^2(\nabla\phi\cdot\nabla\psi)}{1-\kappa^2|\nabla\phi|^2}\,\nabla\phi\right)^2+\beta\kappa^2\left(\frac{{\nabla\phi}\cdot{\nabla\psi}}{1-\kappa^2{|\nabla\phi|}^2}\right)^2}}.\lb{2.34}
\eea
These new dyonic matter equations appear complicated. 
In the limiting case of the first Born--Infeld model \cite{B1,B2,B3,B4}, with $\kappa=0$, however, they reduce themselves into
\bea
\nabla\cdot\left(\nabla\phi\sqrt{\frac{1-\beta|\nabla\psi|^2}{1-\beta|\nabla\phi|^2}}\right)&=&\rho_e,\lb{2.35}\\
\nabla\cdot\left(\nabla\psi\sqrt{\frac{1-\beta|\nabla\phi|^2}{1-\beta|\nabla\psi|^2}}\right)&=&\rho_m,\lb{2.36}
\eea
which are the Euler--Lagrange equations of the action functional
\be
{\cal A}(\phi,\psi)=\int \left(\frac1\beta\left[1-\sqrt{1-\beta|\nabla\phi|^2}\sqrt{1-\beta|\nabla\psi|^2}\right]+\rho_e\phi+\rho_m\psi\right)\,\dd x.
\ee
As a comparison, we note that in this case the Lagrangian action density of the Born--Infeld theory \eq{x2.6} with $\kappa=0$ is
\be\lb{2.38}
{\cal L}=\frac1\beta(1-{\cal R})=\frac1\beta\left(1-\sqrt{\frac{1-\beta|\nabla\phi|^2}{1-\beta|\nabla\psi|^2}}\right),
\ee
in view of  \eq{2.34}. Consequently, the associated Hamiltonian \eq{2.21} becomes
\be\lb{2.39}
{\cal H}=\frac{{\bf E}^2}{\cal R}-{\cal L}
=\frac1\beta\left(\frac{1-\beta^2|\nabla\phi|^2|\nabla\psi|^2}{\sqrt{1-\beta|\nabla\phi|^2}\sqrt{1-\beta|\nabla\psi|^2}}-1\right),
\ee
using \eq{2.30}, \eq{2.34}, and \eq{2.38}. This quantity stays non-negative since it can be examined that
\be
\frac{1-st}{\sqrt{1-s}\sqrt{1-t}}\geq 1,\quad s,t\in[0,1).
\ee
As another limiting case, we consider the classical Born--Infeld model \cite{B3,B4}  situation with $\beta=\kappa^2$ so that the system of the equations \eq{2.32}--\eq{2.34} becomes
\bea
&&\nabla\cdot\left(\frac{{\cal R}_0}{1-\beta|\nabla\phi|^2}\nabla\phi+\frac{\beta(\nabla\phi\cdot\nabla\psi)}{{\cal R}_0}\left[\nabla\psi+\frac{\beta(\nabla\phi\cdot\nabla\psi)}{1-\beta|\nabla\phi|^2}\nabla\phi\right]\right)=\rho_e,\lb{x2.26}\\
&&\nabla\cdot\left(\frac1{{\cal R}_0}\left[(1-\beta|\nabla\phi|^2)\nabla\psi+\beta(\nabla\phi\cdot\nabla\psi)\nabla\phi\right]\right)=\rho_m,\lb{x2.27}\\
&&{\cal R}_0=\sqrt{1-\beta|\nabla\phi|^2-\beta|\nabla\psi|^2+\beta^2 (\nabla\phi\times\nabla\psi)^2}.\lb{x2.28}
\eea

We now study how the  solutions of these dyonic matter equations may be constructed explicitly. Interestingly, although the governing equations \eq{2.32}--\eq{2.34} are derived from
exploring \eq{2.19}, their solutions may be obtained from exploiting \eq{2.18}, on the other hand. In fact, the equations in \eq{2.31} indicate that the fields $\bf D$ and $\bf B$ are
determined by {\em prescribing} the electric and magnetic charge densities, $\rho_e$ and $\rho_m$, respectively. With this fact in mind, we get from \eq{2.18} the equations
\bea
{\bf D}^2&=&(f'(s))^2\left({\bf E}^2+\kappa^2(2+\kappa^2{\bf B}^2)({\bf E}\cdot{\bf B})^2\right), \lb{y2.29}  \\
{\bf D}\cdot{\bf B}&=&f'(s)(1+\kappa^2{\bf B}^2)({\bf E}\cdot{\bf B}).
\eea
These equations are nonlinear and nonhomogeneous equations in the unknowns ${\bf E}^2$ and ${\bf E}\cdot{\bf B}$ which may be solved to determine the quantity $s$ given in
\eq{2.8}. 
Note that, we sometimes replace \eq{y2.29} by \eq{2.18} in the above system to solve for $\bf E$, ${\bf E}^2$, and ${\bf E}\cdot{\bf B}$.
Using this result in \eq{2.18}--\eq{2.19}, we find $\bf E$ and $\bf H$. Therefore the electric and magnetic potentials $\phi$ and $\psi$ given in \eq{2.30} are
obtained.

To illustrate the methodology of this construction, assume for simplicity that the dyonic matter is that of a point charge source,
namely, $\rho_e$ and $\rho_m$ in \eq{2.31} are given by the Dirac distributions concentrated at the origin where a dyonic point-charge source resides.
Specifically, in such a situation, the equation \eq{2.31} assumes the form
\be\lb{3.3}
\nabla\cdot{\bf D}=4\pi q \delta({\bf x}),\quad \nabla\cdot{\bf B}=4\pi g\delta({\bf x}),
\ee
where $q$ and $g$ are electric and magnetic charges, respectively, which may be taken to be positive for definiteness. Solving \eq{3.3}, we see that the nontrivial radial components of
${\bf D}$ and $\bf B$ are simply given by
\be\lb{3.4}
D^r=\frac q{r^2},\quad B^r=\frac g{r^2}.
\ee
With \eq{3.3}--\eq{3.4} and the nontrivial radial components obtained for $\bf E$ and $\bf H$ in \cite{Yang2} in light of the method described,   we can integrate \eq{2.30} to get  
a radially symmetric solution
to the general equations \eq{2.32}--\eq{2.34}, fulfilling the normalized asymptotic condition $\phi,\psi\to0$ as $r\to\infty$, which reduces into a solution to \eq{2.35}--\eq{2.36}
when $\kappa=0$ and a solution to  \eq{x2.26}--\eq{x2.28}
when $\beta=\kappa^2$. These explicit solutions are of independent interests and importance \cite{B3,B4,Yang1,Yang2}.

It should be noted that the formalism of the generalized electrodynamics \eq{2.8} requires  $f'(s)$ stay non-vanishing. This condition is usually well observed due to 
the underlying coupling properties of specific models, such as what are seen in the classical Born--Infeld theory in which the radical root operation imposes a
natural cutoff range for $s$ similar to that in special relativity, and in the polynomial model \eq{4} for which all the coefficients are all positive so that no cutoff is present, and in
the exponential model \cite{H1,H2}
\be
f(s)=\frac1\beta(\e^{\beta s}-1),\quad\beta>0,
\ee
for which $f'(s)>0$ automatically follows.
 In general, it is clear that the condition $f'(0)=1$ ensures $f'(s)\neq0$ in all weak field situations, $|s|\ll1$.

\section{Static dyonic matter equations in general and a no-go example}
\setcounter{equation}{0}

In this section, we use the method formulated in the previous section to come up with a derivation of the dyonic matter equations of the nonlinear electrodynamics in its most general setting. We then present an
example to show that a dyonic solution in the full space may not exist. In other words, combining the existence
result in the previous section and the nonexistence result of this section, we conclude that whether a solution exists for the generalized Born--Infeld dyonic matter
equations depends on the underlying specific nonlinearity of the theory imposed.

To start, we use \eq{2.23}--\eq{2.24} to represent the quantity $s$ in \eq{2.8} as
\bea\lb{x3.1}
2s&=&{\bf E}^2-\frac1{(f'(s))^2}\left({\bf H}+\frac{\kappa^2({\bf E}\cdot{\bf H})}{1-\kappa^2{\bf E}^2}{\bf E}\right)^2+\kappa^2\left(\frac{{\bf E}\cdot{\bf H}}{f'(s)(1-\kappa^2{\bf E}^2)}
\right)^2\nn\\
&=&{\bf E}^2-\frac1{(f'(s))^2}\left({\bf H}^2+\frac{\kappa^2({\bf E}\cdot{\bf H})^2}{1-\kappa^2{\bf E}^2}\right).
\eea
This is an implicit equation relating $s$ to $\bf E$ and $\bf H$ which are determined by a pair of scalar fields $\phi$ and $\psi$ through \eq{2.30}.
 Therefore,  we may rewrite such a
relation as
\be\lb{x3.2}
s=\Om (\nabla \phi,\nabla\psi).
\ee
On the other hand, in view of \eq{2.23} and \eq{2.24}, we can express the electric displacement field ${\bf D}$ given in  \eq{2.18} in terms of $\bf  E$ and $\bf H$:
\be
{\bf D}=f'(s){\bf E}+\frac{\kappa^2({\bf E}\cdot{\bf H})}{f'(s)(1-\kappa^2{\bf E}^2 )}\left({\bf H}+\frac{\kappa^2({\bf E}\cdot{\bf H})}{1-\kappa^2{\bf E}^2}\,{\bf E}\right).\lb{x3.3}
\ee
Thus, inserting \eq{2.30} into \eq{2.24} and \eq{x3.3}, we see that the system \eq{2.31} becomes
\bea
&&\nabla\cdot\left(f'(s)\nabla\phi+\frac{\kappa^2(\nabla\phi\cdot\nabla\psi)}{f'(s)(1-\kappa^2|\nabla\phi|^2 )}\left[\nabla\psi+\frac{\kappa^2(\nabla\phi\cdot\nabla\psi)}{1-\kappa^2|\nabla\phi|^2}\,\nabla\phi\right]
\right)=\rho_e,\lb{x3.4}\\
&&\nabla\cdot\left(\frac1{f'(s)}\left[\nabla\psi+\frac{\kappa^2(\nabla\phi\cdot\nabla\psi)}{1-\kappa^2|\nabla\phi|^2}\nabla\phi\right]\right)=\rho_m,\lb{x3.5}
\eea
where $s$ is given by \eq{x3.2}. These equations appear too complicated to solve in their general setting, although it is clearly seen that the system consisting of
the equations \eq{2.32}--\eq{2.34} is contained in \eq{x3.4}--\eq{x3.5} as a limiting case.

As a nonexistence example, consider the $\arcsin$-model proposed by Kruglov \cite{K3,K4} given by
\be\lb{x3.6}
f(s)=\frac1\beta\arcsin (\beta s),\quad \beta>0.
\ee
Thus \eq{x3.1} becomes
\be\lb{x3.7}
2s={\bf E}^2 -(1-[\beta s]^2)\left({\bf H}^2+\frac{\kappa^2({\bf E}\cdot{\bf H})^2}{1-\kappa^2{\bf E}^2}\right).
\ee
Solving \eq{x3.7}, we find
\be\lb{x3.8}
s=\frac{1-\kappa^2{\bf E}^2+\sqrt{(1-\kappa^2{\bf E}^2)^2+\beta^2 ({\bf H}^2-\kappa^2|{\bf E}\times{\bf H}|^2)({\bf H}^2-{\bf E}^2+\kappa^2[{\bf E}^4
-|{\bf E}\times{\bf H}|^2])}}{\beta^2 ({\bf H}^2-\kappa^2|{\bf E}\times{\bf H}|^2)},
\ee
where $\bf E$ and $\bf H$ are given by \eq{2.30} in  terms of the scalar fields $\phi$ and $\psi$ which leads to the determination of \eq{x3.2}.

We now study the existence and nonexistence of a solution to \eq{x3.4}--\eq{x3.5} for the model \eq{x3.6}.

Inserting \eq{3.4} into \eq{2.18} and applying \eq{x3.6}, we obtain the equation that determines the nontrivial radial component $E^r$ of the electric field $\bf E$:
\be\lb{x3.11}
D^r \sqrt{1-\beta^2 s^2}=E^r (1+\kappa^2[B^r]^2), \quad s=\frac12([E^r]^2-[B^r]^2)+\frac{\kappa^2}2 (B^r)^2 (E^r)^2.
\ee
From \eq{x3.11}, we have
\be\lb{xx3.12}
[E^r]^4+\frac{2(2(1+\kappa^2[B^r]^2)-\beta^2[B^r]^2[D^r]^2)}{\beta^2 [D^r]^2 (1+\kappa^2 [B^r]^2) }[E^r]^2+\frac{\beta^2[B^r]^4-4}{\beta^2(1+\kappa^2 [B^r]^2)^2 }=0.
\ee
The discriminant of this quadratic equation in the updated variable $\xi=[E^r]^2$ reads
\be\lb{xx3.13}
\Delta=\frac4{\beta^2(1+\kappa^2[B^r]^2)^2}\left(\frac{(2(1+\kappa^2 [B^r]^2)-\beta^2 [B^r]^2[D^r]^2)^2}{\beta^2[D^r]^4}+4-\beta^2[B^r]^4\right).
\ee
Inserting \eq{3.4} into \eq{xx3.13}, we have
\be
\Delta=-\frac{16(\beta^2\kappa^2q^2  g^4 -[\beta^2 q^4+\kappa^4 g^4-\beta^2 q^2 g^2]r^4 -2\kappa^2g^2r^8-r^{12})r^4}{\beta^4 q^4(\kappa^2g^2+r^4)^2},
\ee
which becomes negative whenever $r$ is sufficiently small if $g>0$. In other words, the dyonic matter equations \eq{x3.4}--\eq{x3.5} for the $\arcsin$-model \eq{x3.6} subject to
the point-charge sources \eq{3.3} or \eq{3.4} have no compatible solution whatsoever for any $\kappa>0$.

Nevertheless, substituting \eq{3.4} into \eq{x3.11}, we can solve for $E^r$ to obtain the formal result
\be\lb{x3.12}
E^r=\frac{\sqrt{\beta^2 q^2g^2+2r^2\left(\sqrt{r^{12}+2\kappa^2 g^2 r^8+(\beta^2q^4+\kappa^4 g^4-\beta^2 q^2 g^2)r^4-\beta^2\kappa^2 q^2 g^4}-\kappa^2 g^2 r^2-r^6\right)}}{\beta q\sqrt{\kappa^2 g^2+r^4}},
\ee
which is ill defined when $r$ is small if $\kappa>0$, as expected, although it is well defined asymptotically for large $r$:
\be
E^r=\frac q{r^2}\left(1-\frac{\kappa^2 g^2}{r^4}\right)+\mbox{O}(r^{-10}),\quad r\gg1.
\ee
Similarly, the magnetic intensity field is ill defined too when $r$ is small and
\be
H^r=\frac g{r^2}\left(1-\frac{\kappa^2 q^2}{r^4}\right)+\mbox{O}(r^{-10}),\quad r\gg1.
\ee
 If $\kappa=0$, we have shown in \cite{Yang1} that no monopole or dyon would exist. In fact, in all situations, if $q=0$, then $E^r=0$ in
\eq{x3.11}, which renders $s=\frac{(B^r)^2}2$, which cannot stay in the permissible interval  $-1\leq\beta s\leq1$ imposed by the model \eq{x3.6}. In conclusion, the dyonic matter equations
\eq{x3.4}--\eq{x3.5} for the model \eq{x3.6} have no dyonic nor monopole solution in the point-charge situation but only the electric point-charge solution:
\be\lb{x3.15}
E^r=\frac{\sqrt{2} q}{\sqrt{r^4+\sqrt{\beta^2 q^4+r^8}}},
\ee
which may also be obtained by setting $g=0$ in \eq{x3.12}. Thus, with ${\bf E}=\nabla\phi$ or $E^r=\phi'(r)$, we can insert \eq{x3.15} to have
\be\lb{x3.16}
\phi(r)=-\int_r^{\infty} \frac{\sqrt{2} q\,\dd\rho}{\sqrt{\rho^4+\sqrt{\beta^2 q^4+\rho^8}}}=-\frac{\sqrt{2} q}{r_0} h\left(\frac r{r_0}\right),\quad r_0\equiv \beta^{\frac14} q^{\frac12},
\ee
where the dimensionless function
\be
h(x)=\int_x^\infty \frac{\dd y}{\sqrt{y^4+\sqrt{1+y^8}}}
\ee
is one of the generalized hypergeometric functions whose specific form does not concern us here.

In the electrostatic situation, the equations \eq{x3.4}--\eq{x3.5} in the context of the $\arcsin$-model \eq{x3.6} are reduced into a single equation
\be\lb{x3.17}
\nabla\cdot\left(\frac{\nabla\phi}{\sqrt{1-\frac{\beta^2}4 |\nabla\phi|^4}}\right)=\rho_e.
\ee
It is readily verified that \eq{x3.16} is a radially symmetric solution to \eq{x3.17} when $\rho_e({\bf x})=4\pi q \delta({\bf x})$ since $\phi(r)$ given in \eq{x3.16} satisfies
\be
\frac{\phi'(r)}{\sqrt{1-\frac{\beta^2}4[\phi'(r)]^4}}=\frac q{r^2}.
\ee

Subsequently, we will construct some exact solutions to \eq{x3.4}--\eq{x3.5} explicitly when
the charge density distribution functions $\rho_e$ and $\rho_m$ are those of dyonic point charge sources with the nonlinearity function $f(s)$ being quadratic and logarithmic.

\section{Dyonically charged  point sources in quadratic model}
\setcounter{equation}{0}

 In this section, we obtain for the quadratic model the exact solution representing a dyonic point charge and we
establish the roles played by the coupling parameter $\kappa$ in rendering a finite-energy solution and restoration of electromagnetic symmetry. Specifically, we show that,
in the quadratic model, although finite-energy
magnetic monopoles and dyons do not exist when $\kappa=0$, finite-energy dyons, but not monopoles, are present when $\kappa>0$. 

Recall that for the quadratic model \cite{De,K2007,GS,C2015,K2017},  the nonlinearity in \eq{2.8} assumes the simple form 
\be\lb{3.1}
f(s)=s+as^2,
\ee
where $a>0$ is a constant. Using \eq{3.1} in \eq{2.18}, 
 we have
\be\lb{3.2}
{\bf D}=\left(1+a\left[{\bf E}^2-{\bf B}^2+\kappa^2({\bf E}\cdot{\bf B})^2\right]\right)\left({\bf E}+{\kappa^2}({\bf E}\cdot{\bf B})\,{\bf B}\right).
\ee
As before, we will consider a static dyonic point charge source given by \eq{3.3} such that the electric displacement field and magnetic field are radial and described by \eq{3.4}.
Thus, implementing consistency in \eq{3.2}, we see that $\bf E$ is also radially symmetric whose radial component $E^r$ satisfies the cubic equation 
\be\lb{3.5}
\frac{D^r}{1+\kappa^2 (B^r)^2}+a(B^r)^2E^r=\left(1+a\left[1+\kappa^2(B^r)^2\right](E^r)^2\right) E^r,
\ee
whose unique real solution, which is positive in view of \eq{3.4}, can of course be obtained explicitly, although its exact expression is too
cumbersome to present here. Nonetheless, we may content ourselves with an analytic discussion. Indeed, for our purpose, we note that the left-hand side of \eq{3.5} represents a line in
$E^r$ with
a positive vertical intercept and a positive slope and the right-hand side of \eq{3.5} represents a cubic concave up curve in $E^r$ passing through the origin. As a consequence, they have
a unique intersection point in the first quadrant of the coordinate plane which gives rise to the unique positive solution of the equation \eq{3.5}. Furthermore, to get more specific information about $E^r$, we insert \eq{3.4} into \eq{3.5} to obtain
\be\lb{3.6}
qr^6+ag^2(\kappa^2 g^2 +r^4) E^r=(\kappa^2 g^2 +r^4)(a[\kappa^2 g^2 +r^4][E^r]^2+r^4) E^r.
\ee
Thus, from \eq{3.6}, we find the following asymptotic expansions of $E^r$ for $\kappa>0$:
\bea
E^r&=&\frac1\kappa-\frac{(a+\kappa^2)r^4}{2a\kappa^3 g^2}+\frac{qr^6}{2a\kappa^2 g^4}+\mbox{O}(r^8),\quad r\ll1;\lb{3.7}\\
E^r&=&\frac q{r^2}-\frac{q(a[q^2-g^2]+\kappa^2 g^2)}{r^6}+\mbox{O}(r^{-10}),\quad r\gg1.\lb{3.8}
\eea

As a consequence, since the free electric charge density induced from the electric field as a function of the radial variable $r$ reads
\be
\rho_{\mbox{\tiny free}}^e(r)=\frac1{4\pi r^2}\frac{\dd}{\dd r} (r^2 E^r),\quad r>0,
\ee
the total free electric charge of the dyon is
\be
q_{\mbox{\tiny free}}=\int_{\bfR^3}\rho_{\mbox{\tiny free}}^e\,\dd{\bf x}=(r^2 E^r)_{r=0}^{r=\infty}= q,
\ee
which agrees with the prescribed electric charge $q$.

On the other hand, in view of \eq{3.7}, \eq{3.8}, \eq{2.19},  $s$ in \eq{2.8}, and \eq{3.1}, we obtain
\bea
H^r&=&\frac{(a+\kappa^2) q r^4}{a\kappa^3 g^3}+\frac{q^2r^6}{a\kappa^2 g^5}+\mbox{O}(r^8),\quad r\ll1,\\
H^r&=&\frac{g}{r^2}-\frac{g(a[g^2-q^2]+\kappa^2 q^2)}{r^6}+\mbox{O}(r^{-10}),\quad r\gg1.
\eea
Hence, by the same computation, we find the total free magnetic charge of the dyon to be
\be
g_{\mbox{\tiny free}}=(r^2 H^r)_{r=0}^{r=\infty}= g,
\ee
in agreement with the prescribed magnetic charge.

We next consider the energy of a dyonic point source. For this purpose, we insert \eq{3.7} into $s$ in \eq{2.8} to obtain
\bea\lb{3.9}
s&=&\frac{(E^r)^2}2+\frac{1}2(\kappa^2 [E^r]^2-1)(B^r)^2\nn\\
&=&-\frac1{2a}+\frac{qr^2}{2\kappa ag^2}+\frac{\left(a^2[8\kappa^5 g^4-3)-2a\kappa^2+\kappa^4\right)r^4}{8a^2\kappa^4g^2}+\mbox{O}(r^6),\quad r\ll1.
\eea
Inserting \eq{3.9} into \eq{2.21} and applying \eq{3.1}, we have
\be
{\cal H}\lb{3.10}
=\frac q{\kappa r^2}+\frac{a^2(8\kappa^5g^4-3)-2a\kappa^2+2\kappa^4}{4a\kappa^4}+\frac{q(-3\kappa^2+a)r^2}{2a\kappa^3 g^2}+\mbox{O}(r^4),\quad r\ll1.
\ee
On the other hand, in view of \eq{3.8},  $s$ given in \eq{2.8}, and \eq{3.4}, we have
\be\lb{3.11}
s=\frac{q^2-g^2}{2r^4}-\frac{q^2}{r^8}\left(a[q^2-g^2]+\frac{\kappa^2g^2}2\right)+\mbox{O}(r^{-12}),\quad r\gg1.
\ee
Thus, using \eq{3.1}, \eq{3.8}, and \eq{3.11} in \eq{2.21}, along with \eq{3.4}, we obtain
\be\lb{3.12}
{\cal H}=\frac{q^2+g^2}{2r^4}-\frac{\left(a[q^2-g^2]^2+2\kappa^2 q^2 g^2\right)}{4r^8}+\mbox{O}(r^{-12}),\quad r\gg1.
\ee

Combining \eq{3.10} and \eq{3.12}, we see that the total energy of a dyonic point source is indeed a finite quantity,
\be
E=4\pi \int_0^\infty {\cal H}\, r^2\,\dd r<\infty.
\ee

In view of \eq{3.10} and \eq{3.12}, it is seen that there is a local electromagnetic asymmetry but a global electromagnetic symmetry exhibited in the energy density of a dyonic point source
with respect to the roles played by its electric and magnetic charges, $q$ and $g$, respectively.

The calculation above is only valid for $\kappa>0$. The situation where $\kappa=0$ needs to be treated separately, as done in \cite{Yang1}, which we now consider briefly in the present broader context, for completeness, while viewing $\kappa$ as a `switching' parameter.

Setting $\kappa=0$ in \eq{3.5}, we have
\be\lb{xx4.17}
{D^r}+a(B^r)^2E^r=\left(1+a(E^r)^2\right) E^r.
\ee
Inserting \eq{3.4} into \eq{xx4.17}, we may obtain $E^r$ explicitly which is again too complicated to state here. Instead, we write its asymptotic expressions as follows:
\bea
E^r&=&\frac{g}{r^2}+\frac{(q-g)r^2}{2a g^2}+\mbox{O}(r^6),\quad r\ll1,\lb{xx4.18}\\
E^r&=&\frac{q}{r^2}+\frac{aq}{r^6}(g^2-q^2)+\mbox{O}(r^{-8}),\quad r\gg1.\lb{xx4.19}
\eea
Besides, in view of \eq{xx4.18}, \eq{xx4.19}, \eq{2.19}, $s$ in \eq{2.8}, and \eq{3.1}, we obtain
\bea
H^r&=&\frac{q}{r^2}+\frac{q(g-q)r^2}{2ag^3}+\mbox{O}(r^6),\quad r\ll1,\\
H^r&=&\frac{g}{r^2}+\frac{ag(q^2-g^2)}{r^6}+\mbox{O}(r^{-10}),\quad r\gg1.
\eea
These expressions indicate that, near $r=0$,  the electric field  behaves like a  magnetic field,  and vice versa, in leading orders, although they do appear to be purely electric and magnetic
asymptotically near infinity. As a consequence, we see immediately that the free electric charge and free  magnetic charge induced from $E^r$ and $H^r$ are
\be
q_{\mbox{\tiny free}}=q-g,\quad g_{\mbox{\tiny free}}=g-q,
\ee
respectively, which is an alternative  indicator by {\em finite quantities} that the total energy of such a dyon is necessarily divergent \cite{Yang1}.

In fact, in view of \eq{2.21}, \eq{3.4}, \eq{3.1},  \eq{xx4.18}, and \eq{xx4.19}, we get
\bea
{\cal H}&=&\frac{qg}{r^4}+\frac{(q-g)^2}{4ag^2}+\mbox{O}(r^4),\quad r\ll1,\lb{xx4.23}\\
{\cal H}&=&\frac{q^2+g^2}{2r^4}+\mbox{O}(r^{-8}),\quad r\gg1,
\eea
for the Hamiltonian density of the dyon.
This expression clearly indicates that the energy  of the dyon with $q,g>0$ diverges at $r=0$ as anticipated.

\section{Dyonic point charges in logarithmic model}
\setcounter{equation}{0}

In this section, we consider the logarithmic
model as a companion case for comparison and obtain an exact dyonic point-charge solution of finite energy as well when $\kappa>0$. We also show that when $\kappa=0$, although
finite-energy electric and magnetic point charges are still present, indicating the same electromagnetic symmetry as in the classical Born--Infeld model, a dyon of finite energy is excluded
or turned off. As a by-product, we show that, when $\kappa>0$, fine electromagnetic structures of the dyon indicate local asymmetry near the point-charge source and global symmetry away from the source.

Recall that the well-studied logarithmic Born--Infeld nonlinear electrodynamics model \cite{Soleng,Fe,AM,Gaete,K6} is defined by
\be\lb{x4.1}
f(s)=-\frac1\beta \ln\left(1-\beta s\right),\quad \beta>0.
\ee
Inserting \eq{x4.1} into \eq{2.18} and using $s$ in  \eq{2.8}, we have
\be\lb{x4.2}
{\bf D}\left(1-\frac\beta2({\bf E}^2-{\bf B}^2)-\frac{\beta\kappa^2}2({\bf E}\cdot{\bf B})^2\right)={\bf E}+{\kappa^2}({\bf E}\cdot{\bf B})\,{\bf B}.
\ee
In the dyonic point charge situation where the radial components of $\bf D$ and $\bf B$ are as given in \eq{3.4}, we can solve for the nontrivial radial component of $\bf E$ in \eq{x4.2} as before to obtain
\be
E^r=\frac{\sqrt{\kappa^2 g^2+r^4}\sqrt{\beta^2 q^2 g^2+(2\beta q^2+\kappa^2 g^2)r^4+r^8}-(\kappa^2 g^2+r^4)r^2}{\beta q(\kappa^2 g^2 +r^4)}.\lb{x4.3}
\ee
This clumsy looking expression actually enjoys rather simple asymptotic properties of our concern, which are
\bea
E^r&=&\frac1{\kappa}-\frac{r^2}{\beta q}+\frac{(\beta q^2 [2\kappa^2-\beta] +\kappa^4 g^2)r^4}{2\beta^2 \kappa^3 q^2 g^2}+\mbox{O}(r^8),\quad r\ll1,\lb{x4.4}\\
E^r&=&\frac q{r^2}-\frac{q}{r^6}\left(\frac{\beta(q^2-g^2)}2+\kappa^2 g^2\right)+\mbox{O}(r^{-10}),\quad r\gg1.\lb{x4.5}
\eea
Thus, we see that the electric field is finite near the point source and the presence of the parameter $\kappa$ is essential. Moreover, near infinity, in leading order the electric field follows a Coulomb
law of that of a pure electric charge as if no magnetism is present but exhibits its magnetic charge component only in higher-order terms.

We may insert the result \eq{x4.3} into \eq{2.19} to obtain the nontrivial radial component $H^r$ of the magnetic intensity field $\bf H$ whose explicit expression appears complicated and
thus omitted here. Instead, like \eq{x4.4} and \eq{x4.5}, we content ourselves with listing the asymptotic expressions for $H^r$ below:
\bea
H^r&=&\frac{2 r^2}{\beta g}+\frac{q(\beta-2\kappa^2) r^4}{\beta\kappa^3 g^3}+\mbox{O}(r^6),\quad r\ll1,\lb{x4.6}\\
H^r&=&\frac{g}{r^2}-\frac g{r^6}\left(\frac{\beta(g^2-q^2)}2+\kappa^2 q^2\right)+\mbox{O}(r^{-10}),\quad r\gg1.\lb{x4.7}
\eea
As a consequence of \eq{x4.4}--\eq{x4.7}, we obtain the same conclusions for the free electric and magnetic charges as those for the quadratic model, $q_{\mbox{\tiny free}}=q$ and 
$g_{\mbox{\tiny free}}=g$. It is interesting to note that, in leading terms, the induced electric field and magnetic intensity field depend on the prescribed electric and magnetic charges
only, both locally near the center of the dyonic charge source and asymptotically near infinity.

To estimate the dyonic energy, we use \eq{2.21} to get the asymptotic properties of the Hamiltonian density
\bea
{\cal H}&=&\frac q{\kappa r^2}+\frac1\beta\left(\ln\frac{\kappa g^2}{q r^2}-1\right)+\mbox{O}(r^2),\quad r\ll1,\lb{x4.8}\\
{\cal H}&=&\frac{(q^2+g^2)}{2r^4}-\frac{(\beta[q^2-g^2]^2+4\kappa^2 q^2g^2)}{8r^8}+\mbox{O}(r^{-12}),\quad r\gg1.\lb{x4.9}
\eea
These results lead to the finiteness of the total energy of a dyonic point-charge source. As a by-product, we observe that  there is a local asymmetry near the center of the dyonic
charge sources between
the electric and magnetic sectors demonstrated through the induced electric and magnetic fields as well as its energy density profile, and that electromagnetic symmetry prevails again
asymptotically away from the charge sources.

It will be of interest to derive the governing equations describing a static dyonic matter distribution. To this end, we insert \eq{x4.1} into \eq{x3.1} to get
\be\lb{x5.10}
2s={\bf E}^2-(1-\beta s)^2\left({\bf H}^2+\frac{\kappa^2({\bf E}\cdot{\bf H})^2}{1-\kappa^2{\bf E}^2}\right).
\ee
Solving for $s$ in \eq{x5.10}, we obtain
\be\lb{x5.11}
s=\frac{\kappa^2{\bf E}^2+\beta{\bf H}^2-\beta\kappa^2|{\bf E}\times{\bf H}|^2+\sqrt{1-\kappa^2{\bf E}^2}\sqrt{\beta(2-\beta{\bf E}^2)(\kappa^2|{\bf E}\times{\bf H}|^2-{\bf H}^2)-\kappa^2{\bf E}^2 +1}-1}{\beta^2({\bf H}^2-\kappa^2|{\bf E}\times{\bf H}|^2)}.
\ee
Using \eq{2.30} in \eq{x5.11}, we can determine \eq{x3.2},  and then derive the governing equations \eq{x3.4}--\eq{x3.5}. Here we omit the details but only note that the dyonic matter
equations beyond the classical Born--Infeld model \eq{x2.6} are generally much more complicated since inserting \eq{x2.6} into \eq{x3.1} gives us
\be
s=\frac{{\bf E}^2(1-\kappa^2{\bf E}^2)+\kappa^2|{\bf E}\times {\bf H}|^2-{\bf H}^2}{2(1-\kappa^2{\bf E}^2) -\beta({\bf H}^2-\kappa^2|{\bf E}\times{\bf H}|^2)},
\ee
which is relatively much simpler, in contrast with \eq{x3.8} and \eq{x5.11}, for example.

Setting $\kappa=0$ in \eq{x5.11}, we have
\be\lb{x5.13}
s=\frac1\beta-\frac{2-\beta{\bf E}^2}{\beta(1+\sqrt{1-\beta{\bf H}^2(2-\beta {\bf E}^2)})}.
\ee
In view of \eq{x4.1} and \eq{x5.13}, we have
\be\lb{x5.14}
f'(s)=\frac{1+\sqrt{1-\beta{\bf H}^2(2-\beta {\bf E}^2)}}{2-\beta{\bf E}^2}.
\ee
Inserting \eq{x5.14} into \eq{x3.4} and \eq{x3.5} (with $\kappa=0$) and using \eq{2.30}, we arrive at
\bea
&&\nabla\cdot\left(\frac{1+\sqrt{1-\beta|\nabla\psi|^2(2-\beta |\nabla\phi|^2)}}{2-\beta|\nabla\phi|^2}\,\nabla\phi\right)=\rho_e,\lb{x5.15}\\
&&\nabla\cdot\left(\frac{2-\beta|\nabla\phi|^2}{1+\sqrt{1-\beta|\nabla\psi|^2(2-\beta |\nabla\phi|^2)}}\, \nabla\psi\right)=\rho_m,\lb{x5.16}
\eea
which is another new set of dyonic matter equations.
In the point-charge source case with \eq{3.3} and \eq{3.4}, we may set $\kappa=0$ in \eq{x4.3} to get
\be\lb{5.17}
E^r=\frac{q(2r^4+\beta g^2)}{r^2(r^4+\sqrt{r^8+\beta q^2 (2 r^4 +\beta g^2)})}.
\ee
As a consequence, we get the radial component of the magnetic intensity field as well,
\be\lb{5.18}
H^r=\frac{g(r^4+\sqrt{r^8+\beta q^2(2 r^4+\beta g^2)})}{r^2(2r^4+\beta g^2)},
\ee
which gives rise to the radial solution to \eq{x5.15} and \eq{x5.16} through $E^r=\phi'(r)$ and $H^r=\psi'(r)$ in the present dyonic point-charge source situation.  In \cite{Yang1}, we showed that
the dyonic energy in this case diverges at $r=0$. Such a property is also clearly exhibited in \eq{5.17} and \eq{5.18} when $g\neq0$. In other words, when $\kappa=0$, a finite-energy
dyonic point-charge source is tuned off, which is turned back on when $\kappa>0$.

\section{Dyonically charged black holes in quadratic and logarithmic models}\lb{sec6}
\setcounter{equation}{0}

In \cite{Yang2},  dyonic point charges of finite energies in the generalized Born--Infeld theory that generate charged black holes with relegated curvature singularities
are explicitly constructed for the classical Born--Infeld model \cite{B3,B4}  and the exponential model \cite{H1,H2}, with ameliorated and removed curvature singularities,
respectively. In this section, we construct dyonically charged black holes with such relegated curvature singularities in the quadratic and logarithmic models, based on the results
obtained in Sections 4--5. We shall see that, although electromagnetism of the generalized Born--Infeld theory does not contribute to the Arnowitt--Deser--Misner (ADM) energy 
as in the Reissner--Nordstr\"{o}m charged black hole situation based on the Maxwell theory, it does to black-hole thermodynamics.

In order to obtain dyonically charged black hole solutions generated from the Einstein equations coupled with the Born--Infeld type nonlinear electrodynamics of the form \eq{2.8},
consider the spacetime line element in the ordered
spherical coordinates $(t,r,\theta,\phi)$ of the Schwarzschild form
\be\lb{4.6}
\dd \tau^2=A(r)\dd t^2-\frac{\dd r^2}{A(r)}-r^2\left(\dd\theta^2+\sin^2\theta\,\dd\phi^2\right).
\ee
In Appendix, we show that the line element \eq{4.6} is the most general static spherically symmetric one for the construction of
a dyonically charged black hole solution in the generalized  Born--Infeld theory.
Within the ansatz \eq{4.6},  it has been shown \cite{Yang2} that the Einstein equations may be reduced  into the single equation
\be
(rA)'=1-8\pi G r^2 {\cal H}(r),
\ee
which directly relates the metric factor $A$ to the energy density of the electromagnetic sector through an integration,
\be\lb{4.13}
A(r)=1-\frac{2GM}r+\frac{8\pi G}r\int_r^\infty {\cal H}(\rho) \rho^2\,\dd\rho,
\ee
where $M$ is an integration constant which may be taken to be positive to represent a mass.
Since the energy of the dyonic point charge is given by
\be\lb{4.14}
E=\int {\cal H}\sqrt{-g}\,\dd r\dd\theta\dd\phi=4\pi \int_0^\infty {\cal H}r^2\,\dd r,
\ee
thus if this quantity is finite, we may rewrite \eq{4.13} as
\be\lb{4.15}
A(r)=1-\frac{2G(M-E)}r-\frac{8\pi G}r\int_0^r {\cal H}(\rho)\,\rho^2\,\dd\rho.
\ee

For the quadratic model \eq{3.1}, the quantity in \eq{4.14} is finite. Hence, inserting \eq{3.10} into \eq{4.15}, we get
\bea\lb{4.16}
&&A(r)=1-\frac{2G(M-E)}r-8\pi G \left(\frac q{\kappa}+\frac{(a^2[8\kappa^5g^4-3]-2a\kappa^2+2\kappa^4)r^2}{12a\kappa^4}+\frac{q(-3\kappa^2+a)r^4}{10a\kappa^3 g^2}\right)\nn\\
&&\quad\quad \quad+\mbox{O}(r^6),\quad r\ll1.
\eea

With \eq{4.16}, we can examine the curvature singularity of the dyonic black hole solution at the mass and charge center, $r=0$. In fact, recall that the usual Kretschmann invariant 
\cite{Henry,MTW} of the metric \eq{4.6} is given by
\be\lb{dyonic5.2}
K=\frac{(r^2 A'')^2+4(rA')^2+4(A-1)^2}{r^4}.
\ee
As a consequence of \eq{4.16}, we see that a finite-energy dyonic black hole shares the same curvature singularity as that of the Schwarzschild black hole, $K\sim r^{-6}$, and that, under the
critical mass-energy condition,
\be
M=E,
\ee
the singularity is relegated to $K\sim r^{-4}$. This feature is known to be also enjoyed by the charged black hole solutions arising in the classical Born--Infeld theory \cite{Yang1,Yang2}.

Furthermore, applying \eq{3.12} in \eq{4.13}, we obtain
\be\lb{4.19}
A(r)=1-\frac{2GM}r+\frac{4\pi G}{r^2}\left(q^2+g^2-\frac{\left(a[q^2-g^2]^2+2\kappa^2 q^2 g^2\right)}{10 r^4}\right)+\mbox{O}(r^{-10}),\quad r\gg1,
\ee
which in leading orders is of the form of the classical Reissner--Nordstr\"{o}m charged black hole metric.

For the logarithmic model \eq{x4.1}, we can insert \eq{x4.8}  into \eq{4.15} to obtain
\be\lb{x6.20}
A(r)=1-\frac{2G(M-E)}r-8\pi G\left(\frac q\kappa+\frac{r^2}{3\beta}\left[\ln\frac{\kappa g^2}q-\frac13-2\ln r\right]\right)+\mbox{O}(r^4),\quad r\ll1,
\ee
which is seen to be slightly more singular at $r=0$ than that of the quadratic model due to the presence of the factor $\ln r$ but significantly improved upon that of the 
Reissner--Nordstr\"{o}m charged black hole again. Furthermore, inserting \eq{x4.9} into \eq{4.13}, we get
\be\lb{x6.21}
A(r)=1-\frac{2GM}r+\frac{4\pi G}{r^2}\left(q^2+g^2 -\frac{(\beta[q^2-g^2]^2+4\kappa^2 q^2g^2)}{20r^4}\right)+\mbox{O}(r^{-10}),\quad r\gg1.
\ee

Recall that, in terms of 
the metric \eq{4.6},  we may use the Brown--York quasilocal energy formula \cite{BY} to compute the ADM energy  \cite{ADM1959,ADM,ADM1962,Carroll,MTW,Wald} of the system in the full space
and confirm that it agrees with the Schwarzschild black hole mass such that
 the electromagnetic energy in the generalized Born--Infeld theory does not contribute to the ADM gravitational energy \cite{Yang2} as in the Reissner--Nordstr\"{o}m charged black hole situation. In contrast, electromagnetism  in the generalized Born--Infeld
theory does contribute to black-hole thermodynamics, as we
now see below.

Consider the dyonically charged black holes generated from the logarithmic model \eq{x4.1} for instance. For simplicity, we assume a supercritical condition, $M<E$, in order to include the Maxwell theory limit,
$E=\infty$. Then \eq{x6.20} implies $A(r)\to\infty$ as $r\to0$. From \eq{x6.21}, we have $A(r)\to1$ as $r\to\infty$. Thus we encounter the possibilities that $A(r)>0$ for all $r>0$,
giving rise to a naked singularity at $r=0$, $A(r)$ vanishes at exactly one spot, say $r_0>0$, leading to the extremal event horizon at $r=r_0$, and $A(r)$ vanishes at multiple spots
among which there is a right-most one, $r_+>0$, rendering the outer event horizon at $r=r_+$ where the black hole resides so that $A'(r_+)\geq0$ holds necessarily, which determines
the Hawking radiation temperature through the formula \cite{FK,MW}
\be\lb{x6.25}
T_{\mbox{\tiny H}}=\frac{A'(r_+)}{4\pi},
\ee
where if $r_+$ is relatively large then it may be solved by setting \eq{x6.21} to zero to arrive at the approximate equation
\be\lb{x6.26}
A(r)=1-\frac{2GM}r+\frac{4\pi G}{r^2}\left(q^2+g^2 -\frac{(\beta[q^2-g^2]^2+4\kappa^2 q^2g^2)}{20r^4}\right)=0.
\ee
Using \eq{x6.26} with $r=r_+$, we have the following dyonic-charge dependent Hawking temperature
\be
T_{\mbox{\tiny H}}=\frac1{2\pi r_+}\left(1-\frac{GM}{r_+}\right)+\frac{G\left(\beta[q^2-g^2]^2+4\kappa^2 q^2 g^2\right)}{5r^7_+},
\ee
by \eq{x6.25}. Although it is unpractical  to determine $r_+$ by \eq{x6.26} explicitly,  this equation gives us the estimate 
\be
r_+>r_{\mbox{\tiny RN}}\equiv GM+\sqrt{(GM)^2-4\pi G(q^2+g^2)},\quad GM^2>4\pi (q^2+g^2),
\ee
where $r_{\mbox{\tiny RN}}$ denotes the classical Reissner--Nordstr\"{o}m black hole radius. Using $r_{\mbox{\tiny RN}}$ to approximate $r_+$, we have the approximation
\be\lb{x6.29}
T_{\mbox{\tiny H}}\approx \frac1{2\pi r_{\mbox{\tiny RN}} }\left(1-\frac{GM}{r_{\mbox{\tiny RN}}}\right)+\frac{G\left(\beta[q^2-g^2]^2+4\kappa^2 q^2 g^2\right)}{5 r_{\mbox{\tiny RN}}^7}.
\ee
In the weak charge limit, $r_{\mbox{\tiny RN}}\approx 2GM=r_{\mbox{\tiny S}}$ (the Schwarzschild radius), the expression \eq{x6.29} is further reduced into
\be\lb{x6.30}
T_{\mbox{\tiny H}}\approx \frac1{8\pi GM}+\frac{\beta(q^2-g^2)^2+4\kappa^2 q^2 g^2}{640\, G^6 M^7},
\ee
where the first term on the right-hand side is the classical Hawking temperature. This result clearly exhibits the dyonic charge correction to the Hawking temperature arising from
the logarithmic Born--Infeld model \eq{x4.1}.  For the quadratic model \eq{3.1}, similar results follow from \eq{4.19}. For example, the expression \eq{x6.30} is now replaced by
the formula
\be
T_{\mbox{\tiny H}}\approx \frac1{8\pi GM}+\frac{a(q^2-g^2)^2+2\kappa^2 q^2 g^2}{320\, G^6 M^7}.
\ee

\section{Characterization by k-essence cosmology and signature of\\ quadratic model}
\setcounter{equation}{0}

In this section,  we show that k-essence cosmology may be
considered in the individual cases of the generalized Born--Infeld formalism
to understand their shared similarities and distinctions. For example, both the quadratic and logarithmic models describe a big-bang universe and have
the same curvature singularity initially in time. On the other hand, we show that they interpolate different cosmic fluid matters and possess different equations of state such that their
associated adiabatic squared
speeds of sound are confined in different physical ranges. More importantly, we shall see that these generalized models may be used to resolve a density-pressure inconsistency issue 
associated with the classical Born--Infeld model that the induced k-essence fluid density is infinite but the pressure remains finite at the big-bang moment. In particular, 
unlike the classical Born--Infeld model and all of its fractional-powered extensions, the quadratic model is shown to
give rise to an equation of state describing a radiation-dominated era for the early evolution of the universe correctly, in addition to the rejection of the model against the presence 
of a finite-energy monopole. Furthermore, we  show that the quadratic model is the unique choice among all the polynomial models that may lead to a k-essence model giving rise to an early-universe radiation-dominated era. Consequently, the features obtained indicate that, with regard to electromagnetic asymmetry (exclusion of monopoles \cite{Yang1}) and symmetry restoration (inclusion of dyons, Section 4) and
k-essence cosmology (onset of radiation-dominated era in the early universe, this section), the quadratic model is uniquely positioned.

Specifically, we have seen  earlier that, in the context of point-charge sources, the quadratic model \eq{3.1} and the logarithmic model \eq{x4.1} behave rather differently from the classical Born--Infeld model 
\eq{x2.6}, although they all give rise to finite-energy electric point charges. A sharp difference arises, however, in the quadratic model, as well as in the general polynomial model \cite{Yang1},
that a finite-energy magnetic point charge is no longer allowed, a distinctive mechanism referred to as electromagnetic asymmetry \cite{Yang1} which is not observed in
the models \eq{x2.6} and \eq{x4.1}. Interesting, we have shown in the preceding sections that the lost electromagnetic symmetry is restored at the dyonic point charge level in the sense that
finite-energy dyonic point charges are now allowed in all three models. Moreover, with regard to acquiring finite-energy electric, magnetic, and dyonic point charges, the models \eq{x2.6}
and \eq{x4.1} are non-distinguishable. In order to understand the distinctions and similarities of the generalized models \eq{3.1} and \eq{x4.1} in comparison with the classical model
\eq{x2.6}, here we consider the cosmological characterizations of these models. We shall focus our attention on the relation between the matter-wave density and pressure arising from 
the k-essence formalism of these models. We will see that, although these models all give rise to big-bang cosmology, they lead to drastically different density-pressure relations: The classical
model \eq{x2.6} renders an infinite density but a {\em finite} pressure at the big-bang moment, which indicates a discrepancy or inconsistency issue, but the generalized models \eq{3.1} and
\eq{x4.1} accommodate both infinite density and pressure at that moment, thereby resolving such a discrepancy, surprisingly.  More note worthily, we show that the classical model leads
to a dust-dominated era in the early universe, which violates the general consensus that the early universe is radiation dominated instead, but the quadratic model gives rise to a
radiation-dominated era correctly,  with explicit precision in the associated equation of state.

In order to study the cosmological expansion of a universe
propelled by a Born--Infeld type k-essence scalar field, we consider the Lagrangian action density of the general form 
 \cite{Bab1,Ba1,Bab2,Ad,An,Ba2,Al,Ru,Ba3,GGY} 
\be\lb{a1}
{\cal L}=f(X)-V(\vp),
\ee
where $X=\frac12\pa_\mu\vp\pa^\mu\vp=\frac12g^{\mu\nu}\pa_\mu\vp \pa_\nu\vp$, $\vp$ being a real-valued scalar field and $g_{\mu\nu}$ the gravitational metric tensor, and $V$ is a potential density
function, aimed to compare the similarities and differences of various nonlinear electrodynamics models discussed. 
We consider an isotropic and homogeneous universe governed by the Robertson--Walker line element
\be\lb{a4}
\dd s^2=\dd t^2-a^2(t)(\dd x^2 +\dd y^2 +\dd z^2),
\ee
expressed
in Cartesian coordinates, $(t,x,y,z)$, where $a(t)>0$ is the scale factor or radius of the universe to be determined. With the notation
 $\dot{a}=\frac{\dd a}{\dd t}$, etc, and assuming that the scalar field is also only time dependent,  the equation of motion of \eq{a1} reads
\be\lb{a6}
\left(a^3 f'(X)\dot{\vp}\right)\dot{}=-a^3 V'(\vp),
\ee
giving rise to the effective energy density $\rho$ and pressure $P$,
\be\lb{5.12}
\rho=\dot{\vp}^2 f'(X)-(f(X)-V(\vp)),\quad P=f(X)-V(\vp),
\ee
and the Einstein equations are reduced into the single Friedmann equation
\be\lb{5.14}
\left(\frac{\dot{a}}a\right)^2=\frac{8\pi G}3\rho=\frac{8\pi G}3\left(2Xf'(X)-f(X)+V(\vp)\right).
\ee
See \cite{Yang1} for details.
A simple but highly relevant special situation is when the potential density function $V$ is constant, say $V_0\geq0$. In this case, the equation \eq{a6} leads to the result
\be\lb{5.16}
f'\left(\frac12\dot{\vp}^2\right)\dot{\vp}=\frac c{a^3},
\ee
where $c$ is taken to be a nonzero constant to avoid triviality, which may be set to be positive for convenience.  Solving \eq{5.16}, we get
\be
X=h\left(\frac c{a^3}\right),
\ee
say.
Hence \eq{5.14} reads
\be\lb{y5.18}
\left(\frac{\dot{a}}a\right)^2=\frac{8\pi G}3\left(2h\left(\frac c{a^3}\right) f'\left( h\left(\frac c{a^3}\right)\right)-f\left(h\left(\frac c{a^3}\right)\right)+V_0\right),
\ee
which is a closed-form equation in $a$ rendering the dynamic evolution of the universe.

We now specialize on the quadratic model \eq{3.1}, by replacing the constant $a$ there with $\alpha$ here in order to avoid a conflict with the standard notation of the scale factor $a=a(t)$ 
given in \eq{a4}, with setting
\be\lb{5.19}
f(X)=X+\alpha X^2,\quad \alpha>0.
\ee
See also \cite{De,GS,NB,JMW,NG}. In this case, the mass density assumes the form
\be\lb{5.20}
\rho=X+3\alpha X^2 +V_0,
\ee
which is positive definite and gives rise to an expanding universe. Solving \eq{5.16} or
\be
\dot{\vp}+\alpha \dot{\vp}^3=\frac c{a^3},
\ee
we find
\be\lb{5.22}
X=\frac12\dot{\vp}^2=\frac{12^{\frac23}\left(\left[\sqrt{12 a^6+81 \alpha c^2}+9\sqrt{\alpha} c\right]^{\frac23}-12^{\frac13} a^2\right)^2}{72\alpha a^2 \left[\sqrt{12 a^6+81 \alpha c^2}+9\sqrt{\alpha} c\right]^{\frac23}}=h\left(\frac c{a^3}\right).
\ee
In view of \eq{5.20} and \eq{5.22}, the Friedmann equation \eq{5.14} appears rather complicated. However, in leading-order approximations, we have
\bea
&&\rho=\frac{3c}4\left(\frac c{\alpha}\right)^{\frac13}\frac1{a^4}-\frac12\left(\frac c\alpha\right)^{\frac23}\frac1{a^2}+\frac1{6\alpha}+V_0+\mbox{O}(a^2),\quad a\ll 1,\lb{5.23}\\
&&\rho=\frac{c^2}{2a^6}+\mbox{O}(a^{-12})+V_0,\quad a\gg1.\lb{5.24}
\eea
 Using \eq{5.23} in \eq{5.14} and the big-bang initial condition $a(0)=0$, we get
\be\lb{5.25}
a(t)\sim t^{\frac12},\quad t\to 0.
\ee
Besides, if $V_0=0$, then using \eq{5.24} in \eq{5.14}, we obtain the power growth law
\be\lb{5.26}
a(t)\sim t^{\frac13},\quad t\to\infty;
\ee
if $V_0>0$, then using \eq{5.24} in \eq{5.14}, we get the familiar exponential growth law
\be\lb{5.27}
a(t)\sim \e^{\sqrt{\frac{8\pi G V_0}3}\,t},\quad t\to\infty,
\ee
unsurprisingly.
Both situations give rise to the same big-bang cosmological expansion scenario, $a(0)=0$, $\dot{a}(t)>0$ for $t>0$, and $a(t)\to\infty$ as $t\to \infty$. Moreover, since the Kretschmann
scalar of the line element \eq{a4} assumes the form
\be\lb{5.28}
K=\frac{3(\dot{a}^4-2a\dot{a}^2\ddot{a}+2a^2\ddot{a}^2)}{2a^4},
\ee
we arrive at the properties
\bea
&& K(t)\sim t^{-4},\quad t\to0; \lb{5.29}\\
&& K(t)\sim t^{-4},\quad t\to\infty, \quad V_0=0; \quad \lim_{t\to\infty}K(t)=\frac{32\pi^2 G^2 V_0^2}3,\quad V_0>0.\lb{5.30}
\eea
The property \eq{5.29} indicates that the big-bang moment $t=0$ is inevitably a curvature singularity.

The discussion here, especially the result \eq{5.27}, suggests that the potential $V(\vp)$ may well be identified with a field-dependent cosmological constant,
$\Lambda=\Lambda(\vp)=8\pi G V(\vp)$, so that \eq{5.12} gives rise to the matter density $\rho_m$ and matter pressure $P_m$ with
\be
\rho=\rho_m+\frac{\Lambda}{8\pi G},\quad P=P_m-\frac{\Lambda}{8\pi G}.
\ee
Therefore, we are led to the expressions
\be\lb{5.32}
\rho_m=2X f'(X)-f(X),\quad P_m=f(X),
\ee
which is a kind of a parametrized form of the equation of state of the cosmic fluid relating $\rho_m$ and $P_m$. Specifically, for the quadratic model \eq{5.19}, we have
\be\lb{5.33}
\rho_m=X+3\alpha X^2,
\ee
by the first equation in \eq{5.32}. Solving this equation, we have
\be\lb{5.34}
X=\frac1{6\alpha}\left(\sqrt{1+12\alpha\rho_m}-1\right).
\ee
Inserting \eq{5.34} into the second equation in \eq{5.32}, we obtain the equation of state
\be\lb{5.35}
P_m=\left(\frac13+\frac4{3(\sqrt{12\alpha\rho_m+1}+1)}\right)\rho_m,
\ee
explicitly. Since
\be
\lim_{t\to0}\rho_m(t)=\infty,\quad\lim_{t\to\infty}\rho_m(t)=0,
\ee
in view of \eq{5.23} and \eq{5.24}, we arrive at the limits
\be\lb{7.37}
\lim_{t\to0} w_m(t)=\frac13,\quad \lim_{t\to\infty} w_m(t)=1,
\ee
where
\be\lb{7.38}
 w_m=\frac{P_m}{\rho_m}=\frac13+\frac4{3(\sqrt{12\alpha\rho_m+1}+1)}.
\ee
This result indicates that the quadratic model monotonically interpolates two classical linear equations of state, radiation-dominated matter with $w_m=\frac13$ and stiff matter with $w_m=1$,
respectively.

As an analog with the classical Newtonian mechanics, the equation of state \eq{5.35} renders us the adiabatic squared speed of sound $c_s^2$ to be
\be
c_s^2=\frac{\dd P_m}{\dd\rho_m}=\frac2{(\sqrt{12\alpha\rho_m+1}+1)^2}\left(1+2\alpha \rho_m+\frac{8\alpha\rho_m+1}{\sqrt{12\alpha\rho_m+1}}\right).
\ee
This is a monotone decreasing function of $\rho_m$ such that
\be
\lim_{\rho_m\to0}c^2_s=1,\quad \lim_{\rho_m\to\infty} c_s^2=\frac13.
\ee
In particular, we have $\frac13<c_s^2<1$ at any time $t>0$, regardless of the value of the positive parameter $\alpha$. Since the permissible range for the adiabatic squared speed of sound is $[0,1)$, we
conclude that the quadratic model \eq{5.19} is of relevance in giving rise to a meaningful cosmic fluid.

We now briefly consider the logarithmic model \eq{x4.1} such that the k-essence nonlinearity is given by
\be\lb{7.41}
f(X)=-\frac1{\beta}\ln(1-\beta X),\quad \beta>0,\quad 0\leq X<\frac1\beta.
\ee
Inserting \eq{7.41} into \eq{5.16}, we have
\be\lb{7.42}
\frac X{(1-\beta X)^2}=\frac{c^2}{2a^6}.
\ee
Solving for $X$ in \eq{7.42} and observing the range of $X$ stated in \eq{7.41}, we get
\bea\lb{7.43}
X&=&h\left(\frac c{a^3}\right)=\frac1{\beta^2 c^2}\left(a^6+\beta c^2- a^3\sqrt{a^6+2\beta c^2}\right)\nn\\
&=&\frac{c^2}{a^6+\beta c^2 +a^3\sqrt{a^6+2\beta c^2}}.
\eea
On the other hand, inserting \eq{7.41} into \eq{5.14}, we find the density of the cosmic fluid defined by the k-essence to be
\be\lb{7.44}
\rho(X)=\frac{2X}{1-\beta X}+\frac1{\beta}\ln(1-\beta X)+V_0.
\ee
It is clear that $\rho'(X)>0$ and 
\be
\rho(0)=V_0,\quad \lim_{X\to\frac1\beta}\rho(X)=\infty,
\ee
corresponding to $a=\infty$ and $a=0$, respectively. In particular, $\rho>V_0\geq0$, which ensures a forever expanding universe. In view of \eq{7.43} and \eq{7.44}, we have
\bea
\rho&=&\frac{\sqrt{2}c}{\sqrt{\beta} a^3}+\frac1\beta\left(\ln\frac{\sqrt{2} a^3}{\sqrt{\beta} c}-1\right)+V_0-\frac{\sqrt{2} a^3}{4\beta^{\frac32} c}+\mbox{O}(a^6),\quad a\ll1,\lb{7.46}\\
\rho&=&\frac{c^2}{2a^6}-\frac{\beta c^4}{8a^{12}}+V_0+\mbox{O}(a^{-18}),\quad a\gg1.\lb{7.47}
\eea
From \eq{7.46} and \eq{5.14}, we see that the big-bang initial condition $a(0)=0$ gives us the behavior
\be
a(t)\sim t^{\frac23},\quad K(t)\sim t^{-4},\quad  t\to0,
\ee
of the scalar factor and the Kretschmann invariant \eq{5.28}, respectively, near the start of time. Furthermore, comparing \eq{7.47} with \eq{5.24}, we see that the behavior \eq{5.30}
holds asymptotically near time infinity. Following \eq{5.32}, we may express the matter density $\rho_m$ and pressure $P_m$ of the logarithmic model \eq{7.41} as 
\be\lb{7.49}
\rho_m=\frac{2X}{1-\beta X}+\frac1{\beta}\ln(1-\beta X),\quad P_m=-\frac1{\beta}\ln(1-\beta X),
\ee
resulting in the equation of state
\be\lb{7.50}
\rho_m=\frac2\beta\left(\e^{\beta P_m}-1\right)-P_m.
\ee
Besides, using \eq{7.43}, we have
\be\lb{7.51}
\lim_{t\to0} X=\frac1\beta,\quad \lim_{t\to\infty}X=0.
\ee
Hence, applying \eq{7.51} to \eq{7.49}, we have
\be
\lim_{t\to0}\rho_m(t)=\infty, \quad \lim_{t\to0}P_m(t)=\infty;\quad \lim_{t\to\infty}\rho_m(t)=0, \quad \lim_{t\to\infty}P_m(t)=0,
\ee
which are all expected in cosmology. In view of \eq{7.50} and L'H\^{o}pital's rule, we obtain
\be\lb{7.53}
\lim_{t\to0} \frac{P_m}{\rho_m}=0,\quad \lim_{t\to\infty} \frac{P_m}{\rho_m}=1,
\ee
indicating that the logarithmic k-essence model interpolates between the dust-matter and stiff-matter models. The adiabatic squared speed of sound is
\be
c^2_s=\frac{\dd P_m}{\dd\rho_m}=\frac1{2\e^{\beta P_m}-1},
\ee
which decreases in $P_m\in[0,\infty)$ and satisfies
\be\lb{7.55}
\lim_{P_m\to0} c_s^2=1,\quad \lim_{P_m\to\infty}c_s^2=0.
\ee
In particular, the range $0<c^2_s<1$ suggests that the logarithmic k-essence model \eq{7.41} could serve as a relevant cosmic fluid model governing an expanding universe.

Finally, it will be instructive to compare these results with those of the classical Born--Infeld model \eq{x2.6} within our formalism, which in
this context gives rise to the corresponding k-essence Lagrangian action density
\eq{a1} with
\be\lb{7.56}
f(X)=\frac1{\beta}\left(1-\sqrt{1-2\beta X}\right).
\ee
Inserting \eq{7.56} into \eq{5.16}, we find
\be\lb{7.57}
X=\frac{c^2}{2(a^6+\beta c^2)},
\ee
which leads to the density function
\bea\lb{7.58}
\rho(X)&=&\frac{2X}{\sqrt{1-2\beta X}}-\frac1{\beta}\left(1-\sqrt{1-2\beta X}\right)+V_0\nn\\
&=&\frac{2X}{\sqrt{1-2\beta X}(1+\sqrt{1-2\beta X})}+V_0,
\eea
implicating an expanding universe. On the other hand, in view of \eq{7.57} and \eq{7.58}, we have 
\bea
\rho&=&\frac c{\sqrt{\beta}a^3}-\frac1\beta+V_0+\frac{a^3}{2c\beta^{\frac32}}+\mbox{O}(a^6),\quad a\ll1,\lb{7.59}\\
\rho&=&\frac{c^2}{2a^6}-\frac{\beta c^4}{8a^{12}}+V_0+\mbox{O}(a^{-18}),\quad a\gg1.\lb{7.60}
\eea
Note that \eq{7.59} is similar to \eq{7.46} but \eq{7.60} is identical to \eq{7.47}. As a result, we have similar time evolution descriptions for various quantities. Besides, we have
the following nonlinear equation of state \cite{Yang1,CGY}:
\be\lb{7.61}
P_m=\frac{\rho_m}{1+\beta\rho_m},
\ee
which renders the limiting behavior \eq{7.53}, the bound $0<c^2_s<1$, and the limits 
\be\lb{7.62}
\lim_{\rho_m\to\infty}c^2_s=0,\quad \lim_{\rho_m\to0} c_s^2=1,
\ee
corresponding to $t=0$ and $t=\infty$, respectively. Note that in \eq{7.62} we use $\rho_m$ to describe the limits of the adiabatic squared speed of sound
instead because by virtue of \eq{7.57}
or \eq{7.61} the matter pressure $P_m$ stays bounded and enjoys the limits
\be
\lim_{t\to0}P_{m}(t)=\frac1{\beta},\quad \lim_{t\to\infty}P_{m}(t)=0.
\ee
In other words, unlike that in the quadratic and logarithmic models, the matter pressure of the classical Born--Infeld model stays bounded even at the big-bang time $t=0$,
although its matter density tends to infinity there. This is
certainly
 a less desirable, physically unnatural or discrepant, feature. 
Therefore, we conclude that, as cosmic fluids, both the quadratic model \eq{5.19} and the logarithmic model \eq{7.41} significantly deviate from the classical
Born--Infeld model \eq{7.56},  in view of k-essence cosmology, in addition to their shared common features. These detailed fine points are summarized as follows.

\begin{enumerate}
\item[(a)] In the context of k-essence cosmology, the classical Born--Infeld model, the quadratic model, and the logarithmic model all give rise to a big-bang scenario for
the evolution of the universe such that at the initial moment the matter density in each model is infinite and in far future it vanishes.

\item[(b)] The matter pressure in the classical Born--Infeld model remains finite at the initial moment despite of the fact that the matter density is infinite then. However, in both
the quadratic and logarithmic models, the matter pressures become infinite initially at the big-bang moment as well, which is consistent with the infinite matter-density property, and thus may be considered physically more natural or agreeing.

\item[(c)] In terms of the equations of state, the classical Born--Infeld model and the logarithmic model both interpolate between the dust-matter and stiff-matter fluids, but the
quadratic model interpolates between the radiation-dominated matter and stiff matter.

\item[(d)] In either the classical Born--Infeld model and the logarithmic model, the adiabatic squared speed of sound is confined in the unit interval $(0,1)$, but in the quadratic model, it
stays more restrictively in $(\frac13,1)$, reflecting the fact that the associated fluid links the radiation-dominated matter and stiff matter ones.

\item[(e)] The three models also share the common property that their equations of state in each case define the associated fluid pressure as a concave down function of the density,
$\frac{\dd^2 P_m}{\dd\rho_m^2}<0$, such that the adiabatic squared speed of sound $c^2_s$ decreases as $\rho_m$ increases. This property is consistent with the classical
Newton--Laplace sound speed formula \cite{KFC,Ev,Po} for fluids.
\end{enumerate}

In Figure \ref{F1}, these points are further illustrated graphically.
\medskip

Regarding the property (b) stated above, the equations of state \eq{5.35}, \eq{7.50}, and \eq{7.61}, for the quadratic, logarithmic, and the classical Born--Infeld models, respectively, clearly
describe the dependence relations between the fluid density and pressure in each case. In particular, why the pressure in the last case stays finite even though the density approaches infinity
as $t\to0$. Mathematically, such a feature is exhibited in the nonlinearity function \eq{7.56} in that when the quantity $X$ given in \eq{7.57} goes to its threshold value $\frac1{2\beta}$ as
$a\to0$ the function $f(X)$ given by \eq{7.56} goes to a finite limiting value. On the other hand, the function $f(X)$ given by \eq{7.41}, for example,  tends to infinity as $X$ goes to
its threshold value $\frac1{\beta}$ as $a\to 0$ as shown in \eq{7.43}. As for the quadratic model \eq{5.19}, the quantity $X$ has no threshold value and $X\to\infty$ as $a\to0$ so that
both fluid density and pressure go to infinity at the big-bang moment. Such a consistency property has also been established in several other models \cite{Yang1}, although it is lacking in
the original Born--Infeld model as observed.

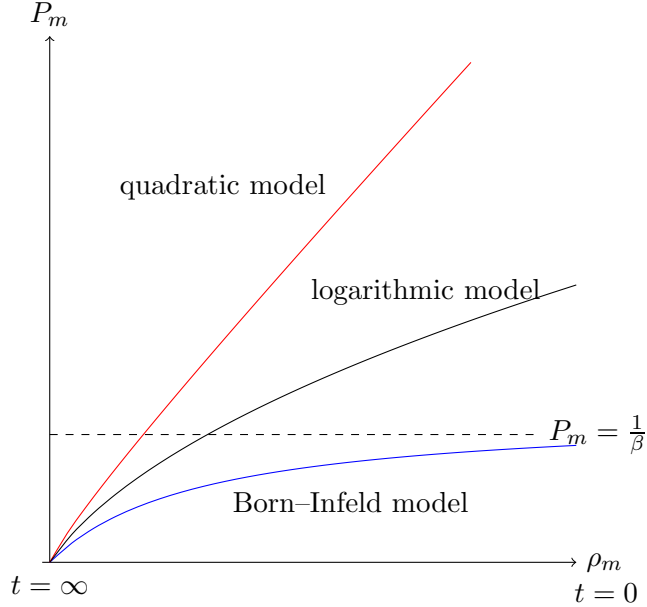
\begin{figure}[H]
\begin{center}
\pgfplotsset{width=7cm,compat=1.8}
\begin{tikzpicture}
  \draw[->] (-0.1, 0) -- (7, 0) node[right] {$\rho_m$};
  \draw[->] (0, 0) -- (0, 7) node[above] {$P_m$};
  \draw[scale=1, domain=0:7, smooth, variable=\x, blue] plot ({\x}, {\x/(1+0.5*\x)});
 \draw[scale=1, domain=0:5.6, smooth, variable=\x, red]  plot ({\x}, {\x*(1+4/(3*(8*(\x)+1)^(0.5)+1))});
 \draw[scale=1, domain=0:7, smooth, variable=\x, black]  plot ({\x}, {((1+3*\x)^(0.5)-1)))});
\draw[-,dashed](0,1.7)--(6.5,1.7) node[right] {$P_m=\frac1\beta$};
\node at (4,0.8) {Born--Infeld model};
\node at (5,3.6) {logarithmic model};
\node at (2.3,5) {quadratic model};
\node at (0,-0.3) {$t=\infty$};
\node at (7.4,-0.4) {$t=0$};
\end{tikzpicture}
\caption{Illustrative plots of the curves of the equations of state of the k-essence cosmic fluids described by the classical Born--Infeld, the logarithmic, and the quadratic models,
as labeled respectively. At the far future corresponding to $t=\infty$ (the coordinate origin on the left), the fluid density $\rho_m$ and pressure $P_m$ of each model both vanish with unit limiting pressure-and-density ratio, $w_m=\frac{P_m}{\rho_m}=1$, and adiabatic squared speed of sound,
$c^2_s=1$. At the initial big-bang moment, $t=0$ (far right), however, these models behave rather differently. For the classical Born--Infeld model, although the fluid density becomes infinite
as $t\to0$, the fluid
pressure stays finite and in fact approaches its limiting value, $\frac1\beta$, as $t\to0$, indicating a discrepancy. In the quadratic and logarithmic models, such a discrepancy is resolved
in the sense that in either of the models the fluid pressure also approaches infinity as $t\to0$.  Moreover, as $t\to0$, the adiabatic squared speeds of sound of the classical Born--Infeld
and logarithmic models both approach zero, implying a dust-matter fluid, but that of the quadratic model tends to $\frac13$, indicating a radiation-dominated-matter fluid.}
\lb{F1}
\end{center}
\end{figure}

It will be of interest to consider various energy conditions \cite{Pen1,Haw,HP,HE,Pen2,Nab,Sen} in our context, which are
\begin{enumerate}
\item[] Weak energy condition: $\rho_m\geq0, \rho_m+P_m\geq0$.

\item[] Dominant energy condition: $\rho_m\geq|P_m|$.

\item[] Strong energy condition: $\rho_m+P_m\geq0, \rho_m+3P_m\geq0$.
\end{enumerate}

From \eq{5.19}, \eq{5.32}, \eq{5.33}, \eq{7.49},  and \eq{7.61},  it is clear that the quadratic,  logarithmic, and classical Born--Infeld models given by \eq{5.19},  \eq{7.41},
and \eq{7.56} satisfy all these energy conditions. 

More generally, we can extend \eq{7.56} to consider the fractional-powered model \cite{Yang1}
\be\lb{7.64}
f(X)=\frac1{\beta}\left(1-\left[1-\frac{\beta X}p\right]^p\right),\quad 0<p<1,
\ee
as in \cite{K1,K2,K5} in the context of nonlinear electrodynamics. Thus, using \eq{5.32}, we obtain the bound $P_m\leq \frac1\beta$ and the equation of state
\be\lb{7.65}
\rho_m=(2p-1)P_m+\frac{2p}\beta\left(\left[1-\beta P_m\right]^{-\left(\frac1p-1\right)}-1\right).
\ee
As a consequence, we have
\be\lb{7.66}
\frac{\dd\rho_m}{\dd P_m}=(2p-1)+\frac{2(1-p)}{(1-\beta P_m)^{\frac1p}}\geq1,\quad P_m<\frac1\beta,
\ee
resulting in the physically desired bounds $0<c_s^2<1$ for all $t>0$ again. Furthermore, combining \eq{7.65} and \eq{7.66}, we arrive at
\be
\lim_{\rho_m\to\infty}P_m=\frac1\beta.
\ee
In other words, the density-pressure inconsistency universally occurs in the fractional-powered model \eq{7.64} as in the classical Born--Infeld model \eq{7.56} as depicted in Figure \ref{F1}.
In particular, we see that the consensus that the early universe is a radiation-dominated era \cite{Wald,Peebles,Weinberg} characterized by the equation of state
\be\lb{7.68}
P_m=\frac13\,\rho_m,
\ee
is {\em violated} by the model \eq{7.64}, as well as by the models \eq{7.41} and \eq{7.56}. In contrast, it is fortunately {\em observed} by the quadratic model \eq{5.19},
as described by \eq{7.37}, or more precisely by \eq{7.38}.  

The above study prompts us to compare the quadratic model in contrast with some other extended models with respect to the preservation of the equation of state \eq{7.68} in the early universe
limit $t\to0$. For this purpose and for convenience, we rewrite \eq{5.16} as
\be\lb{7.69}
f'(X)\sqrt{2X}=\frac c{a^3},\quad c>0,
\ee
so that the dynamic quantity $X=\frac12\dot{\varphi}^2$ depends on the metric factor $a$ monotonically. Now set
\be\lb{7.70}
X_0=\lim_{a\to 0} X=\lim_{t\to0}X.
\ee
We are to consider the limiting ratio
\be\lb{7.71}
w^0_m\equiv \lim_{X\to X_0} w_m(X)=\lim_{X\to X_0}\frac{P_m(X)}{\rho_m(X)}=\lim_{X\to X_0}\frac{f(X)}{2X f'(X)-f(X)},
\ee
in view of \eq{5.32}. 

Indeed, for the polynomial model \cite{Yang1}
\be\lb{7.72}
f(X)=X+\sum_{m=2}^n a_m X^m,\quad a_2,\dots, a_n>0,\quad n\geq2,
\ee
it is clear that $X_0=\infty$,  so that \eq{7.71} renders
\be\lb{7.73}
w_m^0=\lim_{X\to\infty}\frac{X+\sum_{m=2}^n a_m X^m}{X+\sum_{m=2}^n (2m-1)a_m X^m}=\frac1{2n-1}.
\ee
So, to observe a radiation-dominated era with $w^0_m=\frac13$, the unique choice is $n=2$, namely, the quadratic model \eq{5.19}. It is interesting that \eq{7.73} is independent of
the values of the coefficients $a_2,\dots,a_n$ which define the polynomial model \eq{7.72}.

As another example, we consider the exponential model \cite{H1,H2}
\be\lb{7.74}
f(X)=\frac1\beta(\e^{\beta X}-1),\quad\beta >0.
\ee
Inserting \eq{7.74} into \eq{7.69}, we see again that $X_0=\infty$ in \eq{7.70}. Hence, substituting \eq{7.74} into \eq{7.71}, we have
\be\lb{7.75}
w_m^0=\lim_{X\to\infty}\frac{\e^{\beta X}-1}{(2\beta X-1)\e^{\beta X}+1}=0,
\ee
which cannot realize a radiation-dominated era in the early universe, regardless of the value of the parameter $\beta$.

Moreover, it is clear that the exponential model \eq{7.74} is the large-$p$ limit of the fractional-powered model \cite{Yang1}
\be\lb{7.76}
f(X)=\frac1\beta\left(\left[1+\frac{\beta}p X\right]^p-1\right),\quad p\geq1,\quad\beta>0.
\ee
For this model, we also have $X_0=\infty$ so that \eq{7.71} yields
\bea\lb{7.77}
w_m^0(p)&=&\lim_{X\to \infty}\frac{f(X)}{2X f'(X)-f(X)}\nn\\
&=&\lim_{X\to \infty}\frac{1-\left(1+\frac{\beta}p X\right)^{-p}}{2\beta X\left(1+\frac{\beta}pX\right)^{-1}-1+\left(1+\frac{\beta}p X\right)^{-p}}=\frac1{2p-1}.
\eea
Consequently,  \eq{7.75} is the large-$p$ limit of \eq{7.77} as well. Furthermore, the condition of realizing a radiation-dominated era in the early universe, or $w^0_m(p)=\frac13$, singles
out the quadratic model $p=2$ again. Incidentally, when the power $p$ in \eq{7.76} is an integer, $p=2,3,\dots$, the fractional-powered model \eq{7.76} is contained in the polynomial model \eq{7.72}.
 
Thus we conclude that,
through k-essence cosmology, the quadratic model 
distinguishes itself further in giving rise to a correct early-universe equation of state for the wave-matter cosmic fluid it generates.

\section{Conclusions}

The main contributions of this work are the derivation of  the equations of motion governing static dyonic matters, described in terms of two real scalar fields, 
in nonlinear electrodynamics of the Born--Infeld theory type,  construction of exact finite-energy solutions of these equations in the quadratic and logarithmic models
in light of electromagnetic symmetry and asymmetry, subject to dyonic point-charge sources, construction of dyonically charged black holes with relegated curvature singularities
in these models, and cosmological differentiation of the nonlinear models 
under consideration with regard to the associated k-essence dynamics.
Specifically, put in perspective, these new results are summarized and commented on as follows.

\begin{enumerate}

\item[(i)] We have considered the nonlinear electrodynamics of the Born--Infeld theory in its most general formalism and derived its equations of motion
described in terms of two scalar potential fields which govern a static dyonic matter distribution and contain two independent coupling parameters, $\beta$ and
$\kappa$. The first limiting case, $\kappa=0$, corresponds to the first Born--Infeld model arising from special relativity. The second limiting case, $\beta=\kappa^2$,
corresponds to the second Born--Infeld model based on an invariance principle.

\item[(ii)] As important examples, static dyonic matter equations in the classical Born--Infeld model, the quadratic model, and the logarithmic model are presented and their exact
solutions subject to point-charge sources are obtained explicitly. These solutions demonstrate the crucial role played by the coupling parameter $\kappa$ as a switch to turn on and off
the finiteness of the energy carried by a solution.  

\item[(iii)] Of particular interest is the quadratic model situation. In an earlier study, we have seen that a finite-energy electric point charge is accommodated in all nonlinear electrodynamics
models with polynomial type nonlinearity but a magnetic point charge is excluded, within the first Born--Infeld theory context characterized with $\kappa=0$, which is a phenomenon 
referred to as electromagnetic asymmetry. In the present study with $\kappa>0$, the obtained finite-energy dyonic point-charge solutions indicate that electromagnetic symmetry
may be restored in the quadratic model by dyons which accommodate electricity and magnetism simultaneously, although in terms of the fine structures of electromagnetism both
field-wise and energy-wise there is local asymmetry near the center of the dyonic point-charge source but symmetry restoration asymptotically near spatial infinity.
This phenomenon of local asymmetry and asymptotic symmetry of electromagnetism of dyons is further exhibited transparently in the logarithmic model as well.

\item[(iv)] 
 As for the dyonically charged black hole obtained earlier in the classical Born--Infeld model, the charged black holes in the quadratic and logarithmic models have relegated  curvature singularities at the center of the charges, $r=0$,
measured by the Kretschmann invariant, $K$,  which are of the same order as that of the Schwarzschild black hole, $K\sim r^{-6}$, in general, as a result of the finiteness of the
associated electromagnetic energy of the dyonic point-charge sources.
This singularity may further be ameliorated to
$K\sim r^{-4}$ under a critical mass-energy condition, as in the classical Born--Infeld model.

\item[(v)] 
For both the quadratic and logarithmic models, the gravitational metric factors of the dyonically charged black holes are indistinguishable and behave like
the Reissner--Nordstr\"{o}m black hole
 up to the order of $r^{-6}$
for $r\gg1$, although they possess different fine local structures electromagnetically and energetically for $r\ll1$. Up to the order $r^{-2}$, these black holes have the same Brown--York
quasilocal energy within the radial coordinate distance $r\gg1$ such that they possess the same ADM mass to which the electromagnetic energy makes no contribution. However, in these
models, electromagnetism contributes to black-hole thermodynamics through the Hawking radiation.

\item[(vi)] In the context of k-essence cosmology, the quadratic and logarithmic models both give rise to the same big-bang universe scenario and initial curvature singularities, although
as cosmic fluids these models follow
 different equations of state and interpolate different limiting states. In particular, the associated adiabatic squared speeds of sound of the models take different physically meaningful ranges
during the evolution of the universe. 

\item[(vii)] In terms of k-essence cosmology, the quadratic and logarithmic models are shown to distinguish themselves against the classical Born--Infeld model in that the former give rise to a consistent
description for the expansion of the universe in a big-bang scenario such that in these two former models both fluid density and  pressure blow up at the initial moment but the latter
third model renders a
finite limit of the pressure despite  the initial blowup of the fluid density, which signals a physical discrepancy or inconsistency in the classical model in contrast. More importantly and
interestingly, the quadratic model gives rise to a radiation-dominated era in the early universe,  and in contrast, the classical model, and more generally, the fractional-powered model,  the logarithmic
model,  as well as the exponential model,  all  fail to do so, but instead, give rise to a dust-dominated era, rendering another inconsistency against the commonly accepted consensus. In this regard, the quadratic model seems
more favorable and advantageous, in addition to its exclusion of a finite-energy monopole, and stands out uniquely among other polynomial models,
beyond the quadratic model, which all exclude monopoles but give rise
to incorrect pressure-density ratios in the early universe.
\end{enumerate}

\section*{Appendix: metric factors}
\renewcommand{\theequation}{A.\arabic{equation}}
\setcounter{equation}{0}

In this appendix, we elaborate on the metric factor problem associated with dyonically charged black holes generated from the nonlinear electrodynamics of the Born--Infeld type considered in
Section \ref{sec6}. Recall that, with spherical symmetry, the most general static spacetime line element assumes the form \cite{MTW,Wald}
\be\lb{A1}
\dd \tau^2=g_{\mu\nu}\dd x^\mu\dd x^\nu =P(r)\dd t^2-Q(r)\dd r^2-r^2(\dd\theta^2+\sin^2\theta\,\dd\phi^2),
\ee
where $P$ and $Q$ are positive-valued functions of the radial variable $r$ only. It is well known that, in the situations of the vacuum Einstein equations and the Einstein equations coupled
with the Maxwell equations so that the solutions are the Schwarzschild massive black hole and the Reissner--Nordstr\"{o}m charged black hole, respectively, it can be shown that $P$ and $Q$
must satisfy the normalized condition 
\be\lb{A0}
PQ=1.
\ee
 In the context of the generalized nonlinear electrodynamics, it continues to be consistent to assume the condition \eq{A0}. That is, it is consistent to work with the line element \eq{4.6}
as an ansatz. On the other hand, it will be interesting and important to study whether the condition \eq{A0} is dictated, as in the classical situations,  by nonlinear electrodynamics as well.
(This question was raised by an anonymous referee. The author thanks the referee for suggesting this study.) In this appendix, we show, indeed, that \eq{A0} must hold for dyonically charged
black hole solutions as a necessary condition, regardless of the form of the function $f(s)$ in the nonlinear electrodynamics theory \eq{2.8}. In other words, the line element \eq{4.6}
is the most general line element to bring forth static spherically symmetric dyonically charged black hole solutions to the Einstein equations coupled with nonlinear electrodynamics of the Born--Infeld
type.

To proceed, we shall relate the Ricci tensor of the line element 
\eq{A1} to the energy-momentum tensor of the nonlinear electrodynamics defined by \eq{2.8} through the Einstein equations. We see that the governing equations of
electromagnetism impose some constraints to the energy-momentum tensor, hence to the Ricci tensor as well, which inevitably results in \eq{A0}. As a by-product of this formalism, we
also show that electric and magnetic charges arise as integration constants, rather than prescribed quantities, naturally giving rise to dyonically charged black holes. This property is an
interesting aspect for its own sake.

First, with (\ref{A1}), the nontrivial and independent components of the  Ricci tensor are given by the expressions
\bea
R_{00}&=&-\frac{P''}{2Q}+\frac{P'}{4Q}\left(\frac{P'}{P}+\frac{Q'}{Q}\right)-\frac{P'}{rQ},\label{A5}\\
R_{11}&=&\frac{P''}{2P}-\frac{P'}{4P}\left(\frac{P'}P+\frac{Q'}Q\right)-\frac{Q'}{rQ},\label{A6}\\
R_{22}&=&-1+\frac1Q+\frac r{2Q}\left(\frac{P'}P-\frac{Q'}Q\right),\label{A7}\\
R_{33}&=&\sin^2\theta\, R_{22}.\lb{A8}
\eea

Next,  since the electromagnetic tensor now is also radially symmetric \cite{Yang2}, we see that the nontrivial and independent components of $F^{\mu\nu}$ are
\be\lb{A9}
F^{01}=-X,\quad F^{23}=-Y,
\ee
in which $X$ and $Y$ may be regarded as the radial components of the underlying electric and magnetic fields sustained by a dyonic charge, respectively, whose properties will become clearer later. Thus, using $(g_{\mu\nu})=\mbox{diag}(P,-Q,-r^2,-r^2\sin^2\theta)$ and $(g^{\mu\nu})=\mbox{diag}(P^{-1},-Q^{-1},-r^{-2},-r^{-2}\sin^{-2}\theta)$ to lower and raise indices, respectively, we see that the nontrivial and independent components of $F_{\mu\nu}, \tilde{F}_{\mu\nu}$, and $\tilde{F}^{\mu\nu}$ are
 \bea
&& F_{01}=PQ X,\quad F_{23}=-r^4\sin^2\theta\, Y,\lb{A10}\\
&&\tilde{F}_{01}=\sqrt{PQ}\, r^2\sin\theta \,Y,\quad \tilde{F}_{23}=\sqrt{PQ}\,r^2\sin\theta\, X,\lb{A11}\\
&&\tilde{F}^{01}=-\frac{r^2\sin\theta \,Y}{\sqrt{PQ}},\quad \tilde{F}^{23}=\frac{\sqrt{PQ}\,X}{ r^2\sin\theta},\lb{A12}
\eea
respectively. Here $\tilde{F}^{\mu\nu}$ is the Hodge dual of $F^{\mu\nu}$ with respect to the metric $(g_{\mu\nu})$ given by
\be
\tilde{F}^{\mu\nu}=\frac12\varepsilon^{\mu\nu\alpha\beta}F_{\alpha\beta},\quad \varepsilon^{0123}=\frac1{\sqrt{-g}}.
\ee
With these results, we have
\be\lb{A13}
F_{\mu\nu}\tilde{F}^{\mu\nu}=-4r^2\sin\theta\sqrt{PQ}\, XY.
\ee
Moreover, recall that the curved-spacetime Born--Infeld electromagnetic field equations of \eq{2.8}  now read
\bea
\frac1{\sqrt{-g}}\pa_\mu\left(\sqrt{-g}P^{\mu\nu}\right)&=&0,\lb{dyonic4.2}\\
P^{\mu\nu}&=&f'(s)\left(F^{\mu\nu}-\frac{\kappa^2}4[F_{\alpha\beta}\tilde{F}^{\alpha\beta}]\tilde{F}^{\mu\nu}\right),\lb{dyonic4.3}
\eea
where the quantity $s$ in \eq{2.8} and \eq{dyonic4.3} assumes the form
\bea\lb{A16}
s&=&-\frac14 F_{\mu\nu}F^{\mu\nu}+\frac{\kappa^2}{32}\left(F_{\mu\nu}\tilde{F}^{\mu\nu}\right)^2\nn\\
&=&\frac12\left(PQ X^2-r^4\sin^2\theta\, Y^2\right)+\frac{\kappa^2}2 r^4\sin^2\theta\,{PQ}\,(XY)^2,
\eea
by \eq{A9}, \eq{A10}, and \eq{A13}. 
Similarly, in view of \eq{A9}, \eq{A12}, and \eq{A13}, we obtain the nontrivial and independent components of the tensor field $P^{\mu\nu}$ given in \eq{dyonic4.3} to be
\bea
&&P^{01}=-f'(s)\left(1+{\kappa^2} r^4\sin^2\theta \,Y^2\right)X,\lb{A14}\\
&&P^{23}=-f'(s)\left(1-{\kappa^2}{PQ}X^2\right)Y.\lb{A15}
\eea
 On the other hand, from \eq{A12}, we see that the Bianchi identity
\be\lb{A17}
\frac1{\sqrt{-g}}\pa_\mu(\sqrt{-g}\tilde{F}^{\mu\nu})=0,
\ee
has only two independent nontrivial components at $\nu=0$ and $\nu=3$, respectively, which can be reduced into
\bea
&&(r^4\sin^2\theta\, Y)_r=0,\lb{A18}\\
&&X_\theta=0,\lb{A19}
\eea
since $\sqrt{-g}=\sqrt{PQ}\,r^2\sin\theta$. As a consequence of \eq{A19}, we find that $X$ depends on $r$ only. Moreover, within spherical symmetry, the quantity \eq{A16} 
depends on $r$ only. With this fact in mind, we obtain from the $\nu=3$ component of the equation \eq{dyonic4.2} the result
\be
(\sin\theta \,Y)_\theta=0,
\ee
which implies the conclusion
\be\lb{A21}
Y=\frac Z{\sin\theta},
\ee
where $Z$ depends on $r$ only. Hence, in terms of the $r$-dependent functions $X$ and $Z$, we have
\bea
\tilde{F}_{01}&=&\sqrt{PQ}\, r^2 Z,\quad \tilde{F}_{23}=\sqrt{PQ}\,r^2\sin\theta\, X,\lb{A22}\\
\tilde{F}^{01}&=&-\frac{r^2 Z}{\sqrt{PQ}},\quad \tilde{F}^{23}=\frac{\sqrt{PQ}\, X}{r^2\sin\theta},\lb{A23}\\
F_{\mu\nu}\tilde{F}^{\mu\nu}&=&-4r^2\sqrt{PQ}\, XZ,\lb{24}\\
s&=&\frac12\left(PQ X^2-r^4 Z^2\right)+\frac{\kappa^2}2 r^4 PQ\, X^2 Z^2.\lb{A25}
\eea
Inserting \eq{A10} and \eq{A22}--\eq{A25} into 
the energy-momentum tensor given by
\be\lb{4.4}
T_{\mu\nu}=
-f'(s)\left(F_{\mu\alpha}g^{\alpha\beta}F_{\nu\beta}-\frac{\kappa^2}4(F_{\mu'\nu'}\tilde{F}^{\mu'\nu'})F_{\mu\alpha}g^{\alpha\beta}\tilde{F}_{\nu\beta}\right)-g_{\mu\nu}f(s),
\ee
we see that its nontrivial components are
\bea
T_{00}&=&P^2 Q f'(s)\left(1+{\kappa^2} r^4  Z^2\right)X^2-Pf(s),\lb{A26}\\
T_{11}&=&-PQ^2 f'(s)\left(1+{\kappa^2} r^4 Z^2\right)X^2+Qf(s),\lb{A27}\\
T_{22}&=&r^6 f'(s)\left(1-{\kappa^2} PQ  X^2\right) Z^2+r^2 f(s),\lb{A28}\\
T_{33}&=&\sin^2\theta\, T_{22}.\lb{A29}
\eea
Using \eq{A26}--\eq{A29}, we find the trace of the energy-momentum tensor to be
\be\lb{A30}
T=2f'(s)\left(PQ X^2+2{\kappa^2} r^4 PQ X^2 Z^2-r^4 Z^2\right)-4f(s).
\ee

We are now ready to derive the anticipated reduction of the line element \eq{A1}, as stated in \eq{A0}. To this end, we use \eq{A5}, \eq{A6},  \eq{A26}, \eq{A27}, and \eq{A1} to arrive at
\bea
\frac1r\left(\frac{P'}P+\frac{Q'}Q\right)&=&-\frac QP R_{00}-R_{11}\nn\\
&=&8\pi G\left(\frac QP T_{00}+T_{11}\right)-4\pi G\left(\frac QP g_{00}+g_{11}\right)T=0,\lb{A31} 
\eea
in view of the Einstein equations
\be\lb{4.5}
R_{\mu\nu}=-8\pi G\left(T_{\mu\nu}-\frac12 g_{\mu\nu}T\right).
\ee
 Note that the vanishing of the second term on the right-hand side of \eq{A31} is actually independent of the details of $T$. As a consequence of \eq{A31}, we have
\be
(PQ)'=0,
\ee
which leads to $PQ=$ constant. Since we are interested in an asymptotically flat metric, we should impose the condition $P(r),Q(r)\to 1$ as $r\to\infty$. This establishes
\eq{A0} so that we arrive at the line element \eq{4.6}.

As a consequence, with $P=A$ and $Q=\frac1A$, we see that \eq{A5}--\eq{A8} and \eq{A26}--\eq{A30} become 
\be\lb{dyonic4.7}
R_{00}=-\frac{A}{2r^2}(r^2 A')',\quad
R_{11}=\frac{1}{2r^2A}(r^2 A')',\quad
R_{22}=(rA)'-1,\quad
R_{33}=\sin^2\theta\, R_{22},
\ee
 and
\bea
T_{00}&=&Af'(s)\left(1+{\kappa^2} r^4  Z^2\right)X^2-Af(s),\lb{A33}\\
T_{11}&=&-\frac{f'(s)}A\left( 1+{\kappa^2} r^4  Z^2\right)X^2+\frac{f(s)}A,\lb{A34}\\
T_{22}&=&r^6 f'(s)\left( 1-{\kappa^2} X^2\right)Z^2+r^2 f(s),\quad T_{33}=\sin^2\theta\, T_{22},\lb{A35}\\
T&=&2f'(s)\left( X^2+2{\kappa^2} r^4 X^2 Z^2-r^4 Z^2\right)-4f(s),\lb{A36}
\eea
respectively.
Therefore, in view of \eq{dyonic4.7} and \eq{A33}--\eq{A36}, we see that the Einstein equations \eq{4.5} are reduced into
\bea
(r^2 A')'&=&16\pi G r^2\left(r^4 f'(s)(Z^2-\kappa^2[XZ]^2)+f(s)\right),\lb{A37}\\
(rA)'&=&1-8\pi G r^2\left(f'(s)(X^2+\kappa^2 r^4 [XZ]^2)-f(s)\right),\lb{A38}
\eea
which are over-determined.
We now show that \eq{A37} is necessarily contained in \eq{A38}. In fact, in terms of $A, X, Z$, the $\nu=0$ component of \eq{dyonic4.2} becomes
\be\lb{A39}
\left(r^2 f'(s)(1+\kappa^2 r^4 Z^2)X\right)_r=0.
\ee
Thus, applying \eq{A39}, we see that \eq{A37} and \eq{A38} lead to
\be\lb{A40}
(r^2 A')'-r(r A)''=16\pi G r^6 Z f'(s)\left(1-\kappa^2 X^2\right)\left(2Z+\frac{rZ'}2\right),
\ee
in view of
\be\lb{A41}
\frac{\dd s}{\dd r}=
(1+\kappa^2 r^4 Z^2)XX'+r^4(-1+\kappa^2 X^2)ZZ'+2r^3(-1+\kappa^2X^2)Z^2,
\ee
because \eq{A25} is updated into
\be\lb{A42}
s=\frac12\left(X^2-r^4 Z^2\right)+\frac{\kappa^2}2 r^4 X^2 Z^2.
\ee
Since consistency requires $(r^2 A')'-r(rA)''=0$, we are led to imposing the equation 
\be\lb{A43}
rZ'=-4Z
\ee
in \eq{A40}. However, it is clear that \eq{A43} must hold in view of \eq{A18} and \eq{A21}. Hence we have verified that \eq{A37} is contained in \eq{A38}. In other words,
the Einstein equations \eq{4.5} are reduced into the single equation \eq{A38}. Besides, the solution to \eq{A43} reads
\be\lb{A44}
Z=\frac{g}{r^4},
\ee
which represents a magnetic point charge. In fact,  applying \eq{A44} to \eq{A39}, we have
\be\lb{A45}
f'(s)\left(1+\kappa^2 \frac{g^2}{r^4}\right) X=\frac{q}{r^2},
\ee
where the constant $q$ may naturally be interpreted as a prescribed electric charge.  In view of \eq{3.4}, we see that \eq{A45} is the radial version of the constitutive equation
\eq{2.18} with setting
\be
D^r=\frac q{r^2},\quad E^r=X,\quad B^r=r^2 Z=\frac g{r^2},
\ee
so that \eq{A33}--\eq{A36} and \eq{A42} all assume their expected radially symmetric forms in terms of the prescribed electric charge $q$ and magnetic charge $g$, respectively.
Consequently, we arrive at the formulation carried out in Section \ref{sec6} as described.

\medskip
\medskip

{\bf Data availability statement}: The data that supports the findings of this study are
available within the article.

\medskip

The author thanks an anonymous referee whose critical comments and suggestions helped improve the presentation of this work.

\end{document}